\documentclass[a4paper,11pt]{article}

\pdfoutput=1 

\usepackage{jheppub} 

\usepackage[x11names,dvipsnames,svgnames]{xcolor}
\usepackage{hyperref}

\usepackage{tikz}
\usetikzlibrary{patterns,arrows,decorations.pathmorphing,decorations.markings}
\usetikzlibrary{decorations.markings}

\usepackage{youngtab}
\usepackage{soul}

\usepackage{bm,bbm,slashed,empheq,mathrsfs}
\usepackage[normalem]{ulem}

\usepackage[T1]{fontenc} 

\def\rd{{\rm d}}
\def\tr{{\rm Tr}\,}
\def\Tr#1{\langle #1 \rangle}

\def\nn{ \nonumber \\ }
\def\vev#1{\left\langle #1 \right \rangle}

\def\coef#1#2#3{ {\vphantom{A^2}}^{#1}\!{{C}}^{#3}_{#2} }
\def\dcoef#1#2#3{ {\vphantom{A^2}}^{#1}\!{\dot{{C}}}^{#3}_{#2} }

\def\fcs[#1#2]#3{ f^{\phantom{#1#2}#3}_{#1#2} }

\preprint{
	\mbox{}\hfill{} ZU-TH 69/23 \\
	\mbox{}\hfill{} PSI-PR-23-39 
}

\title{\boldmath Two Loop Renormalization of Scalar Theories using a Geometric Approach}

\author[a]{Elizabeth E.~Jenkins,}

\author[a]{Aneesh V.~Manohar,}

\author[a,b,c]{Luca Naterop,}

\author[a]{Julie Pag\`es}

\affiliation[a]{Physics Department 0319,
University of California San Diego,\\ 9500 Gilman Drive, La Jolla, CA 92093-0319, USA}

\affiliation[b]{Physik-Institut, Universit\"at Z\"urich,
Winterthurerstrasse 190, CH-8057 Z\"urich, Switzerland}

\affiliation[c]{Paul Scherrer Institut CH-5232 Villigen PSI, Switzerland}

\emailAdd{ejenkins@ucsd.edu}
\emailAdd{amanohar@ucsd.edu}
\emailAdd{luca.naterop@physik.uzh.ch}
\emailAdd{jcpages@ucsd.edu}

\abstract{
We derive a general formula for two-loop counterterms in Effective Field Theories (EFTs) using a geometric approach. This formula allows the  two-loop results of our previous paper to be applied to a wide range of theories. The two-loop results hold for loop graphs in EFTs where the interaction vertices  contain operators of arbitrarily high dimension, but at most two derivatives. We also extend our previous one-loop result to include operators with an arbitrary number of derivatives, as long as there is at most one derivative acting on each field. The final result for the two-loop counterterms is written in terms of geometric quantities such as the Riemann curvature tensor of the scalar manifold and its covariant derivatives. As applications of our results, we give the two-loop counterterms and renormalization group equations for the $O(n)$ EFT to dimension six, the scalar sector of the Standard Model Effective Field Theory (SMEFT) to dimension six, and chiral perturbation theory to order $p^6$.
}

\allowdisplaybreaks

\begin{document} 
\maketitle
\flushbottom

\section{Introduction}

Observables in quantum field theory (QFT), such as $S$-matrix elements, are unchanged under field redefinitions of the Lagrangian, even though Green functions do change.  Thus, observable quantities are determined in terms of field-redefinition invariants constructed from the Lagrangian. The fields in a QFT take values in a manifold $\mathcal{M}$ such as the curved three-sphere $S^3$ for $SU(2)$ chiral perturbation theory ($\chi$PT).  Field redefinitions correspond to coordinate transformations on $\mathcal{M}$, so field-redefinition invariants are coordinate invariant quantities built out of geometrical objects such as the Riemann curvature tensor of $\mathcal{M}$. The one-loop renormalization of scalar QFTs using a geometrical approach was studied in Ref.~\cite{Alonso:2016oah,Alonso:2015fsp}. The method was generalized to include gauge fields~\cite{Helset:2022pde,Helset:2022tlf} and fermions~\cite{Finn:2020nvn,Gattus:2023gep,Assi:2023zid}. Closely related work by other groups can be found in~\cite{Polyakov:2010pt,Buchalla:2019wsc,Helset:2020yio,Helset:2022pde,Cheung:2022vnd,Cohen:2022uuw,Craig:2023hhp,Alminawi:2023qtf,Alonso:2021rac,Alonso:2022ffe}. In this paper, we restrict our discussion to scalar QFTs.

't~Hooft~\cite{tHooft:1973bhk} computed the one-loop counterterm for a renormalizable scalar field theory containing terms with up to two derivatives in the Lagrangian. The one-loop counterterm formula of 't~Hooft~\cite{tHooft:1973bhk}  requires the scalar kinetic term be canonical, i.e.
\begin{align}
\mathcal{L}_{\text{KE}} &= \frac12 (D_\mu \phi)^i (D^\mu \phi)^i \,,
\label{1.1}
\end{align}
as in a renormalizable QFT. Effective field theories (EFTs) such as the Standard Model Effective Field Theory (SMEFT) or chiral perturbation theory ($\chi$PT) contain non-trivial kinetic terms of the form
\begin{align}
\mathcal{L}_{\text{KE}} &= \frac12 g_{ij}(\phi) (D_\mu \phi)^i (D^\mu \phi)^j \,,
\label{1.2}
\end{align}
where $g_{ij}(\phi)$, the metric on $\mathcal{M}$, in general is non-trivial.
Ref.~\cite{Alonso:2016oah,Alonso:2015fsp} derived a generalization of 't~Hooft's formula that is valid for a general kinetic energy Eq.~\eqref{1.2}. Radiative corrections were given in terms of the Riemann curvature tensor constructed from $g_{ij}$.

In a recent paper~\cite{Jenkins:2023rtg}, referred to as paper I, we derived the two-loop counterterms and anomalous dimensions for a scalar field theory with interactions up to two derivatives and with a canonically normalized kinetic energy term, which is the two-loop version of 't~Hooft's formula. In this paper, we use the geometric approach of refs.~\cite{Alonso:2015fsp,Alonso:2016oah} to extend the results of paper I to an arbitrary scalar theory with interactions up to two derivatives.  Our two-loop result allows for a purely algebraic computation of the two-loop counterterms, i.e.\ we have a universal formula for the counterterms in terms of the Riemann curvature $R_{abcd}$, which is determined from derivatives of the metric $g_{ij}$. 

In this paper, we use the Riemann normal coordinate expansion to compute the fluctuation Lagrangian. The covariant formalism allows for a transformation to a local Cartan frame, where the results of paper I can be used to compute the renormalization counterterms for a general two-derivative scalar theory. The final result is coordinate invariant, so after deriving it, we can use it in any coordinate system. The universal two-loop formula is applied to a number of examples.  It is used to compute the two-loop RGE for the $O(n)$ EFT to dimension six, the two-loop RGE for the Standard Model Effective Field Theory (SMEFT) scalar sector to dimension six, and the two-loop RGE of $SU(n)$ chiral perturbation theory to order $p^6$.
Our geometric approach greatly simplifies the computation --- (a) The corrections are grouped into geometric objects such as the Riemann curvature tensor, which transform covariantly under field redefinitions. (b) The two-derivative cubic interaction in the fluctuation Lagrangian $(D_\mu \eta)(D_\nu \eta) \eta$ vanishes in Riemann normal coordinates, even though it is present under the usual background field method of taking 
$\phi \to \overline \phi + \eta$ and expanding in $\eta$.
Thus, the use of Riemann normal coordinates removes a large number of counterterms that would otherwise have to be computed.  (c) Factorizable diagrams do not contribute to the anomalous dimensions. (d) The field theory for $\eta$ has an $O(n)$ gauge symmetry, noted by 't~Hooft~\cite{tHooft:1973bhk}, which restricts the form of the counterterms, and allows one to determine them without having to compute higher-point amplitudes, even in an EFT. (e) The computations are algebraic, i.e.\ one writes the metric $g_{ij}(\phi)$ as a polynomial in fields, computes the Riemann curvature tensor in terms of derivatives of $g_{ij}$, and takes traces. No Feynman diagrams need to be computed. The basic Feynman graphs give counterterm coefficients which are universal for all EFTs and are tabulated in paper I.

This paper is organized as follows.
In Sec.~\ref{sec:riemann}, we discuss the Riemann normal coordinate expansion, which is used in the derivation of the geometrical version of the two-loop formula.   Sec.~\ref{sec:deriv} then derives the geometrical version of the two-loop results of paper I.  Next, we consider applications to specific scalar field theories. In Secs.~\ref{sec:on}, \ref{sec:smeft} and \ref{sec:chpt}, the algebraic two-loop formula is applied to the $O(n)$ effective field theory to dimension six, SMEFT to dimension six, and chiral perturbation theory to order $p^6$, respectively. Our results for the $O(n)$ EFT agree with known results~\cite{Cao:2021cdt,Cao:2023adc} for $n=1,2$.  Our results for SMEFT agree with known two-loop results for the SM when higher dimension operators are dropped~\cite{Machacek:1984zw,Machacek:1983fi,Machacek:1983tz}, and our results for $\chi$PT agree with Ref.~\cite{Bijnens:1999hw} up to one term discussed in Sec.~\ref{sec:chpt}.  Finally, we present conclusions in Sec.~\ref{sec:conc}.

\section{Riemann Normal Coordinate Expansion}\label{sec:riemann}

The theory we consider is an arbitrary scalar theory with up to two derivatives coupled to external gauge fields. We follow the notation  and analysis of Refs.~\cite{Alonso:2015fsp,Alonso:2016oah}. The Lagrangian is
\begin{align}
\mathcal{L} &= \frac12 g_{ij}(\phi) (D_\mu \phi)^i (D_\mu \phi)^j - V(\phi)
\label{2.1}
\end{align}
where $\phi$ are real scalar fields which take values in a scalar manifold $\mathcal{M}$. Complex scalar fields are written in terms of real scalar fields. The scalar field metric is $g_{ij}(\phi)$, and transforms as a metric tensor under field redefinitions (i.e.\ coordinate transformations of $\mathcal{M}$),
\begin{align}
g_{ij}^\prime &= \left( \frac{\partial \phi^k}{\partial \phi^{\prime\, i}}\right)  \left( \frac{\partial \phi^l}{\partial \phi^{\prime\, j}}\right) g_{kl}\, .
\label{2.2}
\end{align}
Indices on $\mathcal{M}$ are lowered and raised using $g_{ij}$ and its inverse $g^{ij}$.

If the manifold $\mathcal{M}$ on which the scalar fields live has a symmetry, then some of them can be local (gauge) symmetries. In this case, the potential $V(\phi)$ must also be invariant under the gauged symmetries.
The symmetries of $\mathcal{M}$ are generated by Killing vectors $t_\alpha(\phi) $ with vanishing Lie derivative of the metric, $\mathscr{L}_{t_\alpha} g=0$.
They satisfy Killing's equation
\begin{align}
\nabla_i t_{j \alpha} + \nabla_j t_{i \alpha} &=0 \,,
\label{2.3}
\end{align}
where the covariant derivative $\nabla$ uses the metric compatible torsion-free Christoffel connection computed from $g_{ij}$. For symmetries which are gauged, we also require
$\mathscr{L}_{t_\alpha} V=0$ so that the Lagrangian is gauge invariant.

The covariant derivative is
\begin{align}
(D_\mu \phi)^i &= \partial_\mu \phi^i(x) + t^i_\alpha (\phi) A^\alpha_\mu(x)
\label{2.4}
\end{align}
where the gauge fields $A_\mu(x)$ do not depend on $\phi$. The coupling constant and a factor of $i$ has been absorbed into $t_\alpha$. The Killing vectors satisfy the Lie algebra
\begin{align}
\left[ t_\alpha, t_\beta \right] &= \fcs[\alpha \beta]{\gamma} t_\gamma
\label{2.5}
\end{align}
where the l.h.s.\ is a Lie bracket. Explicit expressions for the terms in Eq.~\eqref{2.1} in SMEFT can be found in~\cite{Alonso:2015fsp,Alonso:2016oah,Helset:2020yio,Helset:2022pde}.

Quantum corrections to Eq.~\eqref{2.1} are computed by writing $\phi = \overline \phi + \eta$, where $\overline \phi$ is an external (background) field and $\eta$ is an internal (quantum) field which is integrated over. The one-loop correction is computed from the quadratic term in $\eta$, i.e.\ the second functional derivative of the action. The expansion $\phi = \overline \phi + \eta$ introduces non-covariant terms in the functional derivatives of the action~\cite{Honerkamp:1971sh,Honerkamp:1971xtx}. Under a change of coordinates, a vector $V^i$ transforms as 
\begin{align}
V^{\prime \, i}&=\left( \frac{\partial \phi^{\prime \, i}}{\partial \phi^j}\right) V^j\,.
\label{2.6}
\end{align}
Note that $\phi^i$ is a coordinate which does not transform as a vector. In contrast, $\eta^i = \delta \phi^i$, the variation of $\phi^i$, is a vector, but its derivatives are not~\cite{Honerkamp:1971sh,Honerkamp:1971xtx,Alonso:2015fsp,Alonso:2016oah}.  This introduces non-covariant terms in the functional derivatives and one-loop corrections. However, as pointed out in the introduction, we can make field redefinitions without changing the $S$-matrix. One can make a change of variables to restore manifest covariance of the functional derivatives and the radiative corrections. The idea is to use Riemann normal coordinates to parameterize the quantum fluctuations, which turns functional derivatives into covariant functional derivatives~\cite{Honerkamp:1971sh,Honerkamp:1971xtx,Alonso:2015fsp,Alonso:2016oah}.
%
%
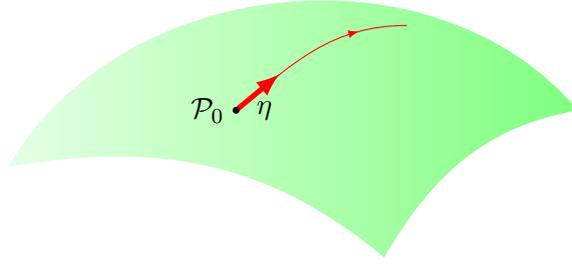
\begin{figure}
\begin{center}
\begin{tikzpicture}[scale=0.75,
mid/.style = {decoration={
    markings,mark=at position 0.753 with {\arrow{latex}}}}]
 \shade[thin, left color=green!10, right color=green!50, draw=none,scale=2]
  (0, 0) to[out=10, in=140] (3.3, -0.8) to [out=60, in=190] (5, 0.5)
    to[out=130, in=60] cycle;
\draw[red,,mid,postaction={decorate}] (4,1) to [out=40, in=180] (7,2.5);

\draw (4,1)+(-0.5,0) node {$\mathcal{P}_0$};
\draw (4,1)+(0.5,0) node {$\eta$};
\draw[-latex,red,line width = 2] (4,1) -- +(0.766,0.642);
\filldraw (4,1) circle (0.05);
 \end{tikzpicture}
\end{center}
\caption{\label{fig:1} Riemann normal coordinates on $\mathcal{M}$. The red curve is a geodesic starting at $\mathcal{P}_0$ with tangent vector $\eta$}
\end{figure}
%
%

Let $\mathcal{P}_0$ be a point on $\mathcal{M}$ with coordinates $\phi_0^i$. This is the point around which we compute the fluctuation Lagrangian. The geodesic $\phi^i(\eta,\lambda)$ is a curve starting at $\mathcal{P}_0$ with tangent vector $\eta_0$ and parameter $\lambda$,
\begin{align}
\phi^i(\eta,0) &= \phi^i_0 , & \frac{\rd \phi^i(\eta,\lambda)}{\rd \lambda}   &= \eta^i (\lambda), & \eta(\lambda=0) &= \eta_0 \, ,
\label{2.7}
\end{align}
which satisfies the geodesic equation $\nabla_\eta \eta=0$, or in coordinates,
\begin{align}
\frac{\rd^2 \phi^i}{\rd \lambda^2} + \Gamma^i_{jk}(\phi) \frac{\rd \phi^j}{\rd \lambda}\frac{\rd \phi^k}{\rd \lambda} &=0\,.
\label{2.8}
\end{align}
The geodesic equation eq.~\eqref{2.8} is homogeneous in $\lambda$, so that $\phi^i(\eta,\lambda) =  \phi^i(s \eta^i, \lambda/s)$ under rescaling by any constant $s$. A point $\mathcal{P}$ with coordinates  $\phi^i(\mathcal{P})$ has Riemann normal coordinates $\eta^i(\mathcal{P})$, where $\phi^i(\mathcal{P} )
= \phi^i(\eta(\mathcal{P}),1)$, i.e.\ the starting velocity at $\mathcal{P}_0$ such that the geodesic reaches $\mathcal{P}$ at $\lambda=1$ is the  Riemann normal coordinate $\eta^i(\mathcal{P})$. The advantage of Riemann normal coordinates is that $\eta(\mathcal{P})$ transforms as a vector at the point $\mathcal{P}_0$ under coordinate transformations.\footnote{Note that it is a vector at $\mathcal{P}_0$, not $\mathcal{P}$.} 
The coordinate $\phi^i$ in Riemann normal coordinates is given by the series expansion
\begin{align}
\phi^i &= \phi^i_0 + \lambda \eta^i - \frac12 \lambda^2\, \Gamma^i_{jk}(\phi_0)\, \eta^j \eta^k - \frac1{3!} \lambda^3\, \Gamma^i_{jkl}(\phi_0)\, \eta^j \eta^k \eta^l
- \frac1{4!} \lambda^4\, \Gamma^i_{jklr}(\phi_0)\, \eta^j \eta^k \eta^l \eta^r+ \ldots
\label{2.9}
\end{align}
where the series coefficients are evaluated by taking successive derivatives of the geodesic equation eq.~\eqref{2.8}, and at each step replacing second derivatives of $\phi^i$ by first derivatives using eq.~\eqref{2.8}.
The higher order terms involve generalized Christoffel symbols~\cite{Eisenhart:1949vn}, which are functions of the standard Christoffel symbols and their derivatives. The background field $\phi_0(x)$ is a function of the spacetime point $x$, so the above procedure gives the quantum field $\eta(x)$ by using eq.~\eqref{2.9} point-by-point in $x$ to parameterize  the fluctuation $\phi(x) - \phi_0(x)$.

The variation of the action is computed by differentiating $S[\phi(\eta,\lambda]$ w.r.t.\ $\lambda$, and evaluating the derivatives at $\lambda=0$. 
Expanding the action in the geodesic fluctuation $\eta$ to quadratic order in $\eta$ yields
\begin{align}
S[\phi + \eta]= S[\phi]+ \frac{\delta S}{\delta \phi^i} \eta^i + \frac12 \left( \frac{\delta^2S}{\delta\phi^i\delta\phi^j} -
\Gamma^k_{ij} \frac{\delta S}{\delta \phi^k}  \right) \eta^i \eta^j +\mathcal{O}(\eta^3)\,,
\label{2.10}
\end{align}
where the terms transform as tensors at $\mathcal{P}_0$. The second functional derivative of $S$ has been replaced by the covariant second functional derivative when using Riemann normal coordinates.

\begin{figure}
\begin{center}
\begin{tikzpicture}[scale=0.6]
\draw (0,0) circle (1);
\draw (-1,0) -- (1,0);
\filldraw (-1,0) circle (0.05);
\filldraw (1,0) circle (0.05);
\draw (0,-2) node {(a)};
\end{tikzpicture}
\hspace{2cm}
\begin{tikzpicture}[scale=0.6]
\draw (-1,0) circle (1);
\draw (1,0) circle (1);
\filldraw (0,0) circle (0.05);
\draw (0,-2) node {(b)};
\end{tikzpicture}
\caption{\label{fig:two} Skeleton graphs for two-loop corrections to the action. There can be an arbitrary number of $\eta\eta$ vertex insertions on the lines.}
\end{center}
\end{figure}
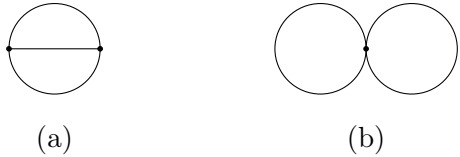

The two-loop contributions to the functional integral are shown in Fig.~\ref{fig:two}.
%
%
%
The graphs involve $\eta^2$, $\eta^3$ and $\eta^4$ interactions, so we need the expansion of the action to fourth order in $\eta$. This can be done efficiently by using a coordinate-free notation, and differentiating the action four times w.r.t.\ the geodesic parameter $\lambda$. We start by computing the derivatives of $D_\mu \phi$, which are needed in the evaluation of derivatives of the action. Its first derivative is
\begin{align}
\nabla_\lambda( D_\mu \phi ) &= \nabla_\lambda (\partial_\mu \phi + t_\alpha A^\alpha_\mu) = \nabla_\lambda (\partial_\mu \phi ) + (\nabla_\lambda t_\alpha) A^\alpha_\mu
\label{2.11}
\end{align}
since the gauge field is independent of the scalar field. The connection is torsion free, $T(X,Y)=\nabla_X Y-\nabla_Y X - [X,Y] = 0$, and $[\partial_\lambda,\partial_\mu]=0$, since partial derivatives commute, so
\begin{align}
\nabla_\lambda (\partial_\mu \phi ) = \nabla_\mu (\partial_\lambda \phi) = \nabla_\mu \eta\,,
\label{2.12}
\end{align}
and
\begin{align}
\nabla_\lambda( D_\mu \phi ) &=  \nabla_\mu  \eta + (\nabla_\lambda t_\alpha) A^\alpha_\mu = \nabla_\mu \eta + (\eta^j \nabla_j t_\alpha) A^\alpha_\mu
\equiv \mathscr{D}_\mu \eta \,,
\label{2.13}
\end{align}
where the covariant derivative $\mathscr{D}_\mu \eta$ was defined in~\cite{Alonso:2015fsp,Alonso:2016oah}. The covariant derivative $\mathscr{D}$ acts on tensors, and can be used to differentiate tensors with an arbitrary number of indices. In coordinates,
\begin{align}
\left(\mathscr{D}_\mu \eta \right)^i &= \left(\partial_\mu \eta^i + \Gamma^i_{kj} \partial_\mu \phi^k \eta^j\right)+A^\beta_\mu \left({t^i_\beta}_{,j} \,   + \Gamma^i_{jk}   \, t^k_\beta \right) \eta^j = \nabla_\mu \eta^i + A^\beta_\mu \eta^j \nabla_j t^i_\beta \nn
&= \left(\partial_\mu \eta^i  +A^\beta_\mu {t^i_\beta}_{,j} \eta^j   \right) + \Gamma^i_{kj} \left(\partial_\mu \phi^k + t^k_\beta A^\beta_\mu \right) \eta^j  = (D_\mu \eta)^i + \Gamma^i_{kj} (D_\mu \phi)^k \eta^j\,.
\label{2.14}
\end{align}
The first form shows that $\mathscr{D}_\mu \eta$ is coordinate invariant, and the second form shows that $\mathscr{D}_\mu \eta$ is gauge invariant. The gauge generators acting on vectors such as $\eta^i$ are ${t^i_\beta}_{,j}$, as shown in~\cite{Alonso:2016oah}.

The geodesic derivative of $ \mathscr{D}_\mu \eta $ is
\begin{align}
\nabla_\lambda \mathscr{D}_\mu \eta &= \nabla_\lambda \left[ \nabla_\mu \eta + (\nabla_\lambda t_\alpha) A^\alpha_\mu \right] = 
 \nabla_\lambda  \nabla_\mu \eta +  (\nabla_\lambda  \nabla_\lambda t_\alpha) A^\alpha_\mu \,.
\label{2.15}
\end{align}
The first term is
\begin{align}
\nabla_\lambda  \nabla_\mu \eta &= [\nabla_\lambda , \nabla_\mu]  \eta +  \nabla_\mu \nabla_\lambda  \eta = R(\eta,\partial_\mu \phi) \eta
\label{2.16}
\end{align}
since $\nabla_\lambda \eta = \nabla_\eta \eta = 0$ by the geodesic equation, and using the definition of the Riemann curvature, $R(X,Y)Z = [\nabla_X,\nabla_Y]Z- \nabla_{[X,Y]} Z$. Eq.~\eqref{2.16} is the geodesic deviation equation,
\begin{align}
\nabla_\lambda  \nabla_\lambda (\partial_\mu \phi) &= R(\eta,\partial_\mu \phi) \eta \,.
\label{2.16a}
\end{align}
We also have
\begin{align}
\nabla_\lambda  \nabla_\lambda t_\alpha &= R(\eta,t_\alpha) \eta
\label{2.17}
\end{align}
which follows from Kostant's equation for a Killing vector  
\begin{align}
\nabla_j \nabla_k\, t^i_\alpha &= R^i{}_{k j l } \, t^l_\alpha \,.
\label{3.19}
\end{align}
Combining eq.~\eqref{2.16} and eq.~\eqref{2.17} gives the gauge invariant form of the geodesic deviation equation,
\begin{align}
\nabla_\lambda ( \mathscr{D}_\mu \eta) &=  R(\eta,D_\mu \phi) \eta \,,
\label{2.18}
\end{align}
for the derivative of $ \mathscr{D}_\mu \eta$.
Eq.~\eqref{2.13} and eq.~\eqref{2.18} are sufficient to compute all higher $\lambda$ derivatives of $D_\mu \phi$. For example, the next derivative is
\begin{align}
\nabla_\lambda \nabla_\lambda ( \mathscr{D}_\mu \eta) &=  \nabla_\lambda  \left[ R(\eta,D_\mu \phi) \eta \right]
= (\nabla_\lambda R) (\eta,D_\mu \phi) \eta + R(\eta, \nabla_\lambda D_\mu \phi) \eta \nn
& = (\eta^i \nabla_i R) (\eta,D_\mu \phi) \eta + R(\eta, \mathscr{D}_\mu \eta ) \eta 
\label{2.19}
\end{align}
since $\nabla_\lambda \eta=0$, and is cubic in $\eta$.

We can now compute the variation of the Lagrangian eq.~\eqref{2.1} to fourth order by repeated application of eq.~\eqref{2.13}, eq.~\eqref{2.18}, and metric compatibility of the connection, $\nabla g=0$. The variations are
\begin{align}
\delta {\cal L} &= g(\mathscr{D}_\mu \eta,D_\mu \phi) - \nabla_\lambda V \,, \nn
\delta^2 {\cal L} &= \frac12 g(R(\eta,D_\mu \phi) \eta ,D_\mu \phi)
+ \frac12  g(\mathscr{D}_\mu \eta,\mathscr{D}_\mu \eta)   - \frac12 \nabla_\lambda  \nabla_\lambda  V \,, \nn
\delta^3 {\cal L}  &= \frac16 g((\nabla_\lambda R)(\eta,D_\mu \phi) \eta ,D_\mu \phi)  + \frac23 g(R(\eta,D_\mu \phi) \eta,\mathscr{D}_\mu \eta)   -\frac16  \nabla_\lambda  \nabla_\lambda  \nabla_\lambda   V \,, \nn
\delta^4 {\cal L}  &= \frac1{24} g((\nabla_\lambda \nabla_\lambda R)(\eta, D_\mu \phi)\eta ,D_\mu \phi) + \frac14 g((\nabla_\lambda R)(\eta, D_\mu \phi)\eta ,\mathscr{D}_\mu \eta) + \frac16 g(R(\eta, \mathscr{D}_\mu \eta)\eta ,\mathscr{D}_\mu \eta)   \nn
& + \frac16 g(R(\eta, D_\mu \phi)\eta , R(\eta, {D}_\mu \phi )\eta )   - \frac1{24} (\nabla_\lambda  \nabla_\lambda  \nabla_\lambda \nabla_\lambda V) \,.
\label{2.21}
\end{align}
where we have included the $1/n!$ prefactor in the $n^{\rm th}$ variation, and used the symmetry of the Riemann tensor
$g(R(A,B)D,C)=g(R(C,D)B,A)$.\footnote{In component notation  $g(R(A,B)D,C) =R_{ijkl} C^i D^j A^k B^l  $. }
The second order term eq.~\eqref{2.21} was given in~\cite[(50)]{Alonso:2015fsp}\cite[(3.42)]{Alonso:2016oah}, and  used to compute the one-loop corrections. Here we need the third and fourth order terms to compute the two-loop corrections. In the next section, we combine eq.~\eqref{2.21} with the result in paper I to  obtain the general two-loop counterterms.

\section{Derivation of the Two-Loop Formula}\label{sec:deriv}

We briefly summarize the results of paper I on the two-loop counterterm formula. The derivation relies on the second, third, and fourth order fluctuations terms having the form
\begin{align}
{\cal L}^{(2)} &= \frac12 (D_\mu \eta)^a (D^\mu \eta)^a + \frac12  X_{ab} \eta^a \eta^b \nn
{\cal L}^{(3)}  &= A_{abc} \eta^a \eta^b \eta^c + A^\mu_{a|bc} (D_\mu \eta)^a \eta^b \eta^c + A^{\mu \nu} _{ab|c} (D_\mu \eta)^a (D_\nu \eta)^b \eta^c \nn
{\cal L}^{(4)}  &=   B_{abcd} \eta^a \eta^b \eta^c \eta^d + B^\mu_{a|bcd} (D_\mu \eta)^a \eta^b \eta^c \eta^d+ B^{\mu \nu} _{ab|cd} (D_\mu \eta)^a (D_\nu \eta)^b \eta^c \eta^d \,.
\label{3.1}
\end{align}
where
\begin{align}
(D_\mu \eta)^a &\equiv \partial_\mu \eta^a + (N_\mu)_{ab} \eta^b \,.
\label{3.2}
\end{align}
and the coefficients $(N_\mu)_{ab}$, $X_{ab}$, etc.\ are functions of the background field, and
satisfy the symmetry relations
\begin{align}
N^\mu_{ab} + N^\mu_{ba} &=0 \,,\nn
A^{\mu}_{a|bc} + A^{\mu}_{b|ca} + A^{\mu}_{c|ab} &= 0\,, \nn
B^{\mu}_{a|bcd} + B^{\mu}_{b|cda} + B^{\mu}_{c|dab} + B^{\mu}_{d|abc} &= 0 \,,
\label{3.3}
\end{align}
i.e.\ the completely symmetric parts of $N^\mu$, $A^\mu$ and $B^{\mu \nu}$ vanish. The coefficients are also symmetric under permutation of $\eta$ indices, and simultaneous permutation of $D_\mu \eta$ indices.  The field-strength tensor is defined by
\begin{align}
Y_{\mu \nu} &= [D_\mu,D_\nu] = \partial_\mu N_\nu - \partial_\nu N_\mu + [N_\mu,N_\nu]\,.
\label{3.4}
\end{align}

The terms in eq.~\eqref{2.21} have the structure of the terms in eq.~\eqref{3.1}. They involve at most two first covariant derivatives of $\eta$, $\mathscr{D}_\mu \eta$.  There are no terms with second derivatives of $\eta$, $\mathscr{D}_\mu \mathscr{D}_\nu \eta$. The cubic term has either $\eta^3$ terms or $\eta \eta \mathscr{D} \eta$ terms; there is no $\eta \mathscr{D}\eta \mathscr{D} \eta$ term with two first derivatives. This feature greatly simplifies the computations required for the two-loop correction. In the extension of 't~Hooft's formula to two loops, we do not need to include the $A^{\mu \nu}_{ab|c}$ interaction. If we had instead used a naive expansion $\phi \to \overline \phi + \eta$, there would in general be such two-derivative terms in the cubic Lagrangian, making the two-loop result far more complicated.
The quartic term has
$\eta^4$, $\eta^3\mathscr{D} \eta$ and $\eta^2 \mathscr{D}_\mu \eta \mathscr{D}_\mu \eta$ terms. The term with two first derivatives in the original Lagrangian eq.~\eqref{2.1} has their Lorentz indices contracted, so the $B^{\mu \nu}_{ab|cd}$ interaction is proportional to $\eta^{\mu \nu}$.

Loop corrections to the action eq.~\eqref{2.1} can be computed using the expansion in quantum fluctuations eq.~\eqref{2.21}. We are free to make field redefinitions, which leave the $S$-matrix invariant. It is not possible to make field redefinitions (coordinate transformations) in the original Lagrangian eq.~\eqref{2.1} to make the metric trivial, $g_{ij}(\phi) \to \delta_{ij}$. The obstruction to finding such a transformation is the non-vanishing of the Riemann curvature tensor $R_{ijkl}(\phi)$, which depends on second derivatives of the metric $g_{ij}$. The 't~Hooft formula is not directly applicable for general Lagrangians such as eq.~\eqref{2.1}, since the formula is only valid for a canonical kinetic term proportional to $\delta_{ij}$.

A big advantage of using Riemann normal coordinates is that the expansion is in powers of $\eta^i$ which transforms like a vector, unlike $\phi^i$ which is a coordinate.   To simplify the radiative corrections, we can go to a local orthonormal (Cartan) frame by introducing vielbeins,
\begin{align}
g_{ij}(\phi) &= e^a_i (\phi) e^b_j(\phi) \delta_{ab}
\label{3.6}
\end{align}
and transform all tensors to the local frame. This frame is a non-coordinate basis, i.e.\ we can make a local choice of tangent vectors satisfying eq.~\eqref{3.6}, but we cannot integrate this to a change of coordinates which makes the metric trivial if the Riemann curvature is non-vanishing. Local frame indices will be denoted by $a,b,c,\ldots$. 
The metric tensor in the local frame is $\delta_{ab}$. The tangent vector $\eta^i$ in the local frame becomes
\begin{align}
\eta^a &= e^a_i(\phi) \eta^i \,,
\label{3.7}
\end{align}
which is a field-redefinition on the quantum field, and does not change the $S$-matrix. 

The variation of the action in the local frame takes the same form as eq.~\eqref{2.21} with coordinate indices $i,j,k,\ldots$ replaced by local frame indices $a,b,c,\ldots$.
The second variation of the action eq.~\eqref{2.21} is
\begin{align}
 \delta^2 {\cal L}  &= \frac12 (\mathscr{D}_\mu  \eta)^a (\mathscr{D}_\mu \eta)^a  + \frac12 R_{abcd} (D_\mu \phi)^a \eta^b \eta^c (D_\mu \phi)^d
  - \frac12 \eta^a \eta^b \nabla_a  \nabla_b  V \,.
\label{3.8}
\end{align}
This second variation has the same form as eq.~\eqref{3.1} with
\begin{align}
X_{ab}  &= -R_{acbd} (D_\mu \phi)^c  (D_\mu \phi)^d  - \nabla_a  \nabla_b  V \,. 
\label{3.9}
\end{align}
The covariant derivative eq.~\eqref{2.14} is
\begin{align}
(\mathscr D_\mu \eta)^a &= \partial_\mu \eta^a + \omega^a_{i b}\, \partial_\mu \phi^i \eta^b + \nabla_b t^a_{\alpha} \eta^b A^\alpha_\mu \,.
\label{3.10}
\end{align}
where $ \omega^a_{i b}$ is the Cartan connection.
Comparing with eq.~\eqref{3.2}, $N^\mu$ is
\begin{align}
N^\mu_{ab} &= \omega_{i a b}\, \partial_\mu \phi^i  + \nabla_b t_{\alpha a}  A^\alpha_\mu \,.
\label{3.11}
\end{align}
where we can lower local frame indices using the metric $\delta_{ab}$. The metric compatible Cartan connection $\omega_{i ab}$ is antisymmetric in $a,b$, as is the derivative of the Killing vector $\nabla_b t_{\alpha a}$, so that $N^\mu_{ab}$ is automatically antisymmetric in $a,b$, as required in eq.~\eqref{3.1}. The commutator of two covariant derivatives eq.~\eqref{3.4} gives~\cite{Alonso:2016oah}
\begin{align}
[Y_{\mu \nu}]^a{}_{b} = [\mathscr D_\mu , \mathscr{D}_\nu] &=  R^a{}_{bcd} (D_\mu \phi)^c (D_\nu \phi)^d + \nabla_b t^a_{\alpha} F^\alpha_{\mu \nu}
\label{3.12}
\end{align}
using properties of the Killing vector. The one-loop counterterm formula of 't~Hooft is
\begin{align}
{\cal L}_{\text{c.t.}}^{(1)}  &=   \frac{1}{16 \pi^2 \epsilon} \left[-\frac14 \tr X_{ab} X_{ba} - \frac{1}{24} \tr (Y_{\mu \nu})_{ab} (Y^{\mu \nu})_{ba} \right]
\label{3.5}
\end{align}
which is only valid if the kinetic term has canonical normalization $\delta_{ab}$, and $(N_\mu)_{ab}$ is antisymmetric. Both conditions are satisfied in the local Cartan frame, so we can use eq.~\eqref{3.5} with $X$ and $Y_{\mu \nu}$ given in eq.~\eqref{3.9} and eq.~\eqref{3.12}.

At first sight, this seems like an involved computation, having to first determine the vielbeins which satisfy eq.~\eqref{3.6}, transform to the local frame, etc. However, the final expression for the counterterm eq.~\eqref{3.5} is coordinate invariant and written in terms of tensor quantities. It can be evaluated in any frame. In particular, it can 
be evaluated in the original coordinate basis --- i.e.\ the entire transformation to the local frame is not actually needed to compute the counterterms; it is only needed to make sure that the formul\ae\ in paper I can be applied. 
For example, the one-loop counterterm in the local frame eq.~\eqref{3.5} becomes
\begin{align}
\mathcal{L} &= \frac{1}{16 \pi^2 \epsilon} \left[ -\frac14  X_{ij} X_{kl}\,  g^{ik} g^{jl} -\frac{1}{24} [Y^{\mu\nu}]^i{}_{j} [Y_{\mu \nu}]^j{}_i \right]
\label{3.14}
\end{align}
in the original coordinate basis, where we have to be careful to raise and lower indices with the metric tensor $g_{ij}$ and its inverse $g^{ij}$. The coordinate basis expressions for $X$ and $Y$ are
\begin{align}
X_{ij}  &= -R_{ikjl} (D_\mu \phi)^k  (D_\mu \phi)^l  -  \nabla_i  \nabla_j  V \,, \nn
[Y_{\mu \nu}]^i{}_j  &=[\mathscr{D}_\mu, \mathscr{D}_\nu ] = R^i{}_{jkl} (D_\mu \phi)^k (D_\nu \phi)^l + \nabla_j t^i_{\alpha} F^\alpha_{\mu \nu}\,,
\label{3.15}
\end{align}
and $D_\mu \phi$ is given in eq.~\eqref{2.3}. Eq.~\eqref{3.15} is the result  derived in ref.~\cite{Alonso:2015fsp,Alonso:2016oah}. Eq.~\eqref{3.15} can be computed using the original Lagrangian, without having to transform to a local frame. The terms have the same form as eq.~\eqref{3.9} and eq.~\eqref{3.12} with local frame indices replaced by coordinate indices.
This method was used to compute the one-loop SMEFT anomalous dimensions in the bosonic sector to dimension eight in ref.~\cite{Helset:2022pde}.

\newpage
For the two-loop correction, we need the cubic plus quartic terms in eq.~\eqref{2.21}, from which we can read off the expansion coefficients in eq.~\eqref{3.1}
\begin{align}
A_{abc} &=  -\frac16  \nabla_{(a}  \nabla_b  \nabla_{c)}  V -\frac{1}{18} \left(  \nabla_a R_{bdce} +  \nabla_b R_{cdae} +   \nabla_c R_{adbe}   \right) (D_\mu \phi)^d  (D^\mu \phi)^e \nn
A^\mu_{a | bc} &=  \frac13 \left( R_{abcd}  + R_{acbd}  \right) (D^\mu \phi)^d  \nn
A^{\mu \nu}_{ab | c} &= 0 \nn
B_{abcd} &=  - \frac1{24} \nabla_{a}  \nabla_b  \nabla_c \nabla_{d} V  -  \frac{1}{24} \nabla_a \nabla_d R_{becf} (D_\mu \phi)^e (D^\mu \phi)^f \nn
& +  \frac16  R_{eabf}   R_{ecdg} (D_\mu \phi)^f (D^\mu \phi)^g \quad  \text{sym}(abcd)
\nn
B^\mu _{a | bcd} &= \frac{1}{4} \left(  \nabla_d   R_{abce}  \right) (D^\mu \phi)^e  \quad \text{sym}(bcd)   \nn
B^{\mu \nu}_{ab | cd} &=  -\frac1{12}  \eta^{\mu \nu} \left( R_{acbd} + R_{adbc} \right)
\label{3.16}
\end{align}
The notation $\text{sym}(abcd)$ and $\text{sym}(bcd)$ means the expressions have to be completely symmetrized in $abcd$ and in $bcd$, i.e.\ we sum over the $n!$ permutations and divide by $n!$. A remarkable fact is that the coefficients $A^\mu$ and $B^\mu$ automatically satisfy the symmetry relations eq.~\eqref{3.9} using the Bianchi identities $R_{abcd}+R_{bcad}+R_{cabd}=0$ and $\nabla_a R_{bcde} + \nabla_b R_{cade}+ \nabla_c R_{abde}=0$, just like the earlier result that $N^\mu$ was automatically antisymmetric in the local frame. We also find that $A^{\mu \nu}$ automatically vanishes. The above equations are tensorial, and so can be used in the original coordinate basis $i,j,k,\ldots$ being careful about raising and lowering indices using $g_{ij}$.

This derivation has been a lengthy one, but the final result is simple --- Compute the quantities $A,A^\mu,B,B^\mu, B^{\mu \nu}$ using the expressions in eq.~\eqref{3.16} in the original coordinate frame, and substitute into the counterterm expressions eq.~(4.1) and eq.~(4.3) of paper I, contracting upper indices with lower indices and raising and lowering indices using $g_{ij}$.

The counterterms in paper I depend on the tensors $X,Y,A,A^\mu,B,B^\mu,B^{\mu \nu}$ and their covariant derivatives. Multiple covariant derivatives  are given by repeated applications of the covariant derivative $\mathscr{D}$ in eq.~\eqref{2.13} which acts on tensors. $\phi$ is not a tensor, and its first derivative is $D_\mu \phi = \partial_\mu \phi + t_\alpha A^\alpha_\mu$. Since $D_\mu \phi$ is a tensor, subsequent covariant derivatives are given by $\mathscr{D}$. There are two formul\ae\ which can be used to simplify multiple covariant derivatives,
\begin{align}
\mathscr{D}_\mu (D_\nu \phi)^i - \mathscr{D}_\nu (D_\mu \phi)^i &= t^i_\alpha F^\alpha_{\mu \nu}\,, \nn
[\mathscr{D}_\mu, \mathscr{D}_\nu ]^i{}_j &= (Y_{\mu \nu})^i{}_j\,.
\label{3.17}
\end{align}
These can be used to reorder the indices on multiple covariant derivatives of $\phi$ and transform them to the form $\mathscr{D}_\mu (D^\mu \phi)$ which can be removed by a field redefinition. The tensors $X,Y,A,A^\mu,B,B^\mu,B^{\mu \nu}$ also contain the Riemann curvature tensor, the potential $V$, and their derivatives. Their
Lie derivatives w.r.t.\ the Killing vectors $t_\alpha$ vanish. Any tensor $T(\phi)$ with arbitrary number of upper and lower indices and vanishing Lie derivative w.r.t.\ the Killing vectors $t_\alpha$ has covariant derivative
\begin{align}
\mathscr{D}_\mu  T &= (D_\mu \phi)^i\, ( \nabla_i T ) \,.
\label{3.18}
\end{align}
If a tensor $T$ satisfies $\mathscr{L}_{t_\alpha} T=0$, then so does it covariant derivative $\nabla T$, because $[\mathscr{L}_{t_\alpha} , \nabla]=0$ when $t_\alpha$ is a Killing vector, which can be proved from the Kostant formula for a Killing vector eq.~\eqref{3.19}.
Eq.~\eqref{3.18} then gives the covariant derivatives of $A$, $A^\mu$, etc.\ needed to compute the counterterms.

An EFT has an expansion in powers of $1/M$, where $M$ is a high energy scale much larger than the scales in the EFT. The potential $V$ up to terms of order $\phi^4$ is from the renormalizable dimension-four part of the Lagrangian, so the terms up to four derivatives in $V$ power-count as dimension four terms. Note that even terms such as $m^2 \phi^2$  are power-counted as dimension four terms rather than dimension two, because $m$ is a low scale in the EFT. The mass term can be written in terms of the EFT power counting as $(m/M)^2 M^2 \phi^2$, i.e.\ as a dimension two operator with a $m^2/M^2$ suppressed coefficient, and so acts as dimension four. The same argument applies to other terms with dimension $<4$ such as $\phi^3$ interactions.
The curvature $R_{abcd}$ has terms with two derivatives of the metric, or the square of one derivative of the metric. These arise from $(\phi/M)^2$ or the square of $\phi/M$ terms in the metric respectively, so $R_{abcd} \sim 1/M^2$ is a dimension-six object. Similarly, $\nabla R \sim 1/M^3 $ is dimension seven, and $R^2 \sim 1/M^4$ and $\nabla \nabla R \sim 1/M^4$ are dimension eight.

\section{The $O(n)$ EFT}\label{sec:on}

The $O(n)$ EFT is the $O(n)$ model including higher dimension operators. Our formalism applies to Lagrangians with up to two derivatives, so the $O(n)$ Lagrangian has the form
\begin{align}
{\cal L} &= \frac12 A\,  (\partial_\mu \phi \cdot \partial_\mu \phi) + B\,  (\phi \cdot \partial_\mu \phi)(\phi \cdot \partial_\mu \phi) - V
\label{4.1}
\end{align}
where $A$, $B$ and $V$ are functions of $(\phi \cdot \phi)$, and $\phi$ is an $n$-component real scalar field.  Loop corrections will generate higher derivative terms in the Lagrangian, which are implicitly added to eq.~\eqref{4.1}. While these are not included in the formul\ae\ derived earlier, we can still consistently compute radiative corrections using eq.~\eqref{4.1} --- the result is the value of the counterterms in the complete theory when the higher derivative coefficients are set to zero. For example, in SMEFT we can compute the $\beta$-function for the dimension eight $H^4 D^4$ operators  $(D_{\mu} H^{\dag} D_{\nu} H) (D^{\nu} H^{\dag} D^{\mu} H)$ from insertions of dimension-six operators, but not the contribution of the dimension-eight operator to its own $\beta$-function.

One can make a field redefinition,
\begin{align}
\phi^i &= \varphi^i f(\varphi \cdot \varphi)
\label{4.2}
\end{align}
and eliminate the $B$ term in eq.~\eqref{4.1}. We will work to dimension-six using the Lagrangian
\begin{align}
{\cal L} &= \frac12 (\partial_\mu \phi \cdot \partial_\mu \phi) - \Lambda - \frac12 m^2 (\phi \cdot \phi) - \frac{\lambda}{4}  (\phi \cdot \phi)^2
+ C_E (\phi \cdot \phi) (\partial_\mu \phi \cdot \partial_\mu \phi) + C_1  (\phi \cdot \phi)^3 +C_2 (\phi \cdot \partial_\mu \phi)^2 \,,
\label{4.3}
\end{align}
where $\Lambda$ is the cosmological constant,  and $C_E, C_1, C_2$ are coefficients of the dimension-six terms and have mass dimension $-2$. The field redefinition eq.~\eqref{4.2} allows one to remove a linear combination of $C_E$ and $C_2$; the linear combination which cannot be eliminated is $2 C_E - C_2$. We work in the basis where $C_2=0$. The loop corrections generate $C_2$, and we use field redefinitions to eliminate $C_2$ and go back to the $C_2=0$ basis.

The metric and inverse metric are
\begin{align}
g_{ij} &= \delta_{ij} + 2 C_E (\phi \cdot \phi) \,, &
g^{ij} &= \delta_{ij} - 2 C_E (\phi \cdot \phi) \,,
\label{4.4}
\end{align}
the Christoffel symbol is
\begin{align}
\Gamma^i_{jk} &= \delta_{ij} + 2 C_E \left[ \delta_{ik} \phi_j + \delta_{ij} \phi_k - \delta_{jk} \phi_i \right]\,,
\label{4.5}
\end{align}
and the curvature is
\begin{align}
R_{ijkl} &= 4 C_E \left[ \delta_{il} \delta_{jk} - \delta_{ik} \delta_{jl} \right]\,,
\label{4.6}
\end{align}
where the difference between up and down indices for $\Gamma^i_{jk}$ and $R_{ijkl}$ is dimension eight. Including counterterms, the bare Lagrangian is
\begin{align}
{\cal L} &= \frac12  Z_\phi (\partial_\mu \phi \cdot \partial_\mu \phi) -[ \Lambda + \Lambda_{\text{c.t.}}] - \frac12 (m^2 + m^2_{\text{c.t.}} ) (\phi \cdot \phi) - \frac14 \mu^{2\epsilon} Z_\phi^2 (\lambda + \lambda_{\text{c.t.}})  (\phi \cdot \phi)^2 \nn
& + \mu^{2\epsilon}  Z_\phi^2 (C_E +C_{E\text{c.t.}}) (\phi \cdot \phi) (\partial_\mu \phi \cdot \partial_\mu \phi) + \mu^{4\epsilon} Z_\phi^3 (C_1 + C_{1\text{c.t.}})  (\phi \cdot \phi)^3  \,.
\label{4.7}
\end{align}

The explicit expressions for the counterterms are given in Appendix~\ref{app:on}. The anomalous dimensions and 't~Hooft consistency conditions can be computed from the counterterms, using the expressions given in paper I. The anomalous dimensions are (where $\dot C \equiv \mu\, \rd C/\rd \mu$)\footnote{All $\{\ \}_1$ terms should be multiplied by
$1/(16\pi^2)$ and all $\{\ \}_2$ terms by $1/(16 \pi^2)^2$. \label{foot}}
\begin{align}
\dot \Lambda &= \left\{ \frac12 n m^4 \right\}_1  +  \left\{ 0 \right\}_2 
\label{4.14}
\end{align} 
\begin{align}
\dot m^2 &= \left\{2(n+2)\lambda m^2 - 8 n  m^4 C_E\right\}_1  +  \left\{ -10(n+2) \lambda^2 m^2 + \frac{80}{3} (n+2)  \lambda m^4 C_E \right\}_2 
\label{4.15}
\end{align} 
\begin{align}
\dot \lambda &= \left\{2(n+8) \lambda^2 -16 (n+3) \lambda m^2 C_E -24 (n+4) m^2 C_1 \right\}_1 \nn
&   +  \left\{ -12 (3n+14) \lambda^3
+\frac{32}{3}(22n+113) \lambda^2  m^2  C_E+480 (n+4)\lambda m^2 C_1\right\}_2 
\label{4.16}
\end{align} 
\begin{align}
\dot C_E &= \left\{4 (n+2)  \lambda C_E \right\}_1  +  \left\{ -34(n+2) \lambda^2 C_E \right\}_2 
\label{4.18}
\end{align} 
\begin{align}
\dot C_1 &= \left\{20 \lambda^2  C_E + 6 (n+14) \lambda C_1 \right\}_1  +  \left\{-\frac83(23n+259) \lambda^3 C_E
-42(7n+54) \lambda^2 C_1  \right\}_2 
\label{4.17}
\end{align} 
where the subscripts outside the curly braces denote the one-loop and two-loop contributions. The 't~Hooft consistency conditions are satisfied. The anomalous dimensions of the $O(n)$ EFT were compute previously in refs.~\cite{Cao:2021cdt,Cao:2023adc} for $n=1$ and $n=2$. We agree with their results for $n=1$ to two-loop order; they have results to 5 loops. They use a  dimension-six basis different 
from ours for $n=2$. The parameters in eq.~\eqref{4.3} in terms of those of refs.~\cite{Cao:2021cdt,Cao:2023adc} are
\begin{align}
\lambda &= \frac14 g ,& C_E &= \frac14 g  C_4^{(6)} , & C_2 &= -\frac14 g C_4^{(6)} , & C_1 &= \frac{1}{3! \ 3!\ 8} g^2 C_6^{(6)} .
\label{4.20}
\end{align}
Our two-loop results for $n=2$ agree with refs.~\cite{Cao:2021cdt,Cao:2023adc}, which give results up to four loops.\footnote{There is a typo in refs.~\cite{Cao:2021cdt,Cao:2023adc}, where the $\phi^6$ operator is given with a normalization $1/(6!\ 8)$ instead of $1/(3!\ 3!\ 8)$. We thank Jasper Nepveu for informing us of this. With this change, our result differs in the sign of the $45g^2/2$ two-loop term in the $C_6^{(6)} -C_4^{(6)}$ entry of the anomalous dimension matrix. }

We find that the 't~Hooft consistency condition is not satisfied for $Z_\phi$, which implies that the field anomalous dimension is infinite,
\begin{align}
\gamma_\phi &= \left\{ -4 (n-1)  m^2 C_E \right\}_1  + \frac{1}{\epsilon}  \left\{ 4(n-1)(n+2) \lambda m^2 C_E   \right\}_2 +  \left\{   (n+2) \lambda^2 -\frac83(n+2)  \lambda m^2 C_E \right\}_2 
\label{4.19}
\end{align} 
The infinite contribution to $\gamma_\phi$ is proportional to $m^2$ times a dimension-six coefficient, and to $(n-1)$.  The anomalous dimensions eq.~\eqref{4.14}--\eqref{4.17} are finite, even though $\gamma_\phi$ enters in their computation through the $Z_\phi$ factors in eq.~\eqref{4.7}.

\begin{figure}
\begin{center}
\begin{tikzpicture}[scale=0.75,
decoration={
	markings,
	mark=at position 0.55 with {\arrow[scale=1.5]{stealth'}};
}]

\draw[postaction=decorate] (90:1) arc (90:270:1) ;
\draw[postaction=decorate] (-90:1) arc (-90:90:1) ;

\filldraw (-0.1,0.9) rectangle (0.1,1.1);

\filldraw (270:1) circle (0.075);

\draw[postaction=decorate] (90:1) -- +(50:1.5) ;
\draw[postaction=decorate] (90:1)+(130:1.5) -- (90:1) ;

\draw[decorate,decoration={snake}]  (270:1) -- +(270:1.5);

\end{tikzpicture}
\end{center}
\caption{\label{fig:penguin} Penguin graph}
\end{figure}
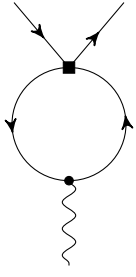

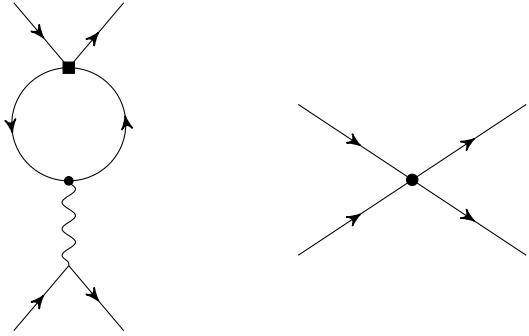
\begin{figure}
\begin{center}
\begin{tikzpicture}[scale=0.75,
decoration={
	markings,
	mark=at position 0.55 with {\arrow[scale=1.5]{stealth'}};
}]

\draw[postaction=decorate] (90:1) arc (90:270:1) ;
\draw[postaction=decorate] (-90:1) arc (-90:90:1) ;

\filldraw (-0.1,0.9) rectangle (0.1,1.1);

\filldraw (270:1) circle (0.075);

\draw[postaction=decorate] (90:1) -- +(50:1.5) ;
\draw[postaction=decorate] (90:1)+(130:1.5) -- (90:1) ;

\draw[decorate,decoration={snake}]  (270:1) -- +(270:1.5);

\draw[postaction=decorate] (270:2.5) -- +(-50:1.5) ;
\draw[postaction=decorate] (270:2.5)+(-130:1.5) -- (270:2.5) ;

\end{tikzpicture}
\hspace{2cm}
\raise1cm\hbox{\begin{tikzpicture}[
decoration={
	markings,
	mark=at position 0.55 with {\arrow[scale=1.5]{stealth'}};
}]

\filldraw (0,0) circle (0.075);
\draw[postaction=decorate]  (-1.5,1) -- (0,0);
\draw[postaction=decorate] (0,0) -- (1.5,1);
\draw[postaction=decorate]  (-1.5,-1) -- (0,0);
\draw[postaction=decorate] (0,0) -- (1.5,-1);
\end{tikzpicture}}
\end{center}
\caption{\label{fig:4} $\psi + \psi \to \psi + \psi$ scattering amplitude.}
\end{figure}

An infinite $\gamma_\phi$ arises because of the use of field redefinitions. Instead of the usual shift $\phi \to \overline \phi + \eta$, we use the expansion in eq.~\eqref{2.9}, which corresponds to a non-linear redefinition of $\eta$, $\eta \to \eta - \Gamma^i_{jk} \eta^j \eta^k/2 + \ldots$. We can understand the origin of infinite field anomalous dimensions from the well-studied case of penguin graphs in the weak interactions. Consider the one-loop graph of
Fig.~\ref{fig:penguin} due to the insertion of a four-fermion operator $ (\overline \psi \gamma_\mu \psi) (\overline \psi \gamma^\mu \psi)$, which leads to a divergence proportional to $(\overline \psi \gamma_\mu \psi) \partial_\nu F^{\mu \nu}$. One can make a field redefinition to replace the $(\overline \psi \gamma_\mu \psi) \partial_\nu F^{\mu \nu}$ by a four-fermion counterterm 
$ (\overline \psi \gamma_\mu \psi) (\overline \psi \gamma^\mu \psi)$. The $\psi + \psi \to \psi + \psi$ scattering amplitude has contributions from the penguin graph and the $\psi^4$ counterterm, as shown in Fig.~\ref{fig:4}, and the sum of the two graphs is finite, and gives a finite $S$-matrix element. However, the three-point correlation function $\vev{\overline \psi \psi A_\mu}$ given by the penguin graph in Fig.~\ref{fig:penguin} is infinite, as there is no longer a $(\overline \psi \gamma_\mu \psi) \partial_\nu F^{\mu \nu}$ counterterm to cancel the divergence. As a result the field $A_\mu$ is not finite, and can have divergent correlation functions. The reason is that the field redefinition to remove the $(\overline \psi \gamma_\mu \psi) \partial_\nu F^{\mu \nu}$ operator is an infinite field redefinition. In our example, the divergent contribution to $\gamma_\phi$ is proportional to $(n-1)$, and so vanishes if there is only a single scalar field, in which case the Riemann curvature vanishes. The coupling constants $m^2$, $\lambda$, etc.\ are related to observable $S$-matrix elements, and so are not affected by field redefinitions.
%
%

%
%
%

An interesting example of an infinite field anomalous dimension was discussed in detail recently~\cite{Bednyakov:2014pia,Herren:2017uxn,Herren:2021yur}. The authors found that the three-loop $\beta$-functions in the SM were divergent, which could be compensated by making an infinite flavor rotation of the fields.

\newpage
\section{SMEFT}\label{sec:smeft}

We apply our two-loop formula to compute the two-loop scalar renormalization in SMEFT to dimension six. We  include the SM Higgs sector, as well as insertions of the dimension six operators
\begin{align}\label{5.0}
& \coef{}{H}{},\ \coef{}{H\Box}{},\ \coef{}{HD}{},\ \coef{}{HG}{},\ \coef{}{HW}{},\ \coef{}{HB}{},\ {\coef{}{HWB}{}},\
 \coef{}{H\widetilde G}{},\ \coef{}{H\widetilde W}{},\ \coef{}{H\widetilde B}{},\ {\coef{}{H\widetilde WB}{}}\,.
\end{align}
and compute their anomalous dimensions from internal scalar loops. The two-loop counterterms are given in App.~\ref{app:smeft}.

The anomalous dimensions due to scalar loops, with subscripts denoting the one-loop and two-loop contributions, are
\begin{align}
\dot \Lambda &= \left\{ \frac{1}{2} m_H^4 \right\}_1 + \left\{ 0\right\}_2 \,,
\label{5.1}
\end{align}
\begin{align}
\dot m_H^2  &=  \left\{ 12 \lambda - 4 m_H^2  C_{H\Box} + 2 m_H^2 C_{HD}   \right\}_1 m_H^2  +
\lambda  \left\{ -60 \lambda + 80m_H^2  C_{H\Box}  - 20 m_H^2  C_{HD} \right\}_2 m_H^2 \,,
\label{5.2}
\end{align}
\begin{align}
\dot \lambda &=  \left\{ 24 \lambda^2  +12 m_H^2 C_H  -32\lambda   m_H^2 C_{H\Box}   +12 \lambda m_H^2 C_{HD}  \right\}_1 \nn
& +
\left\{-312 \lambda^3 -240  \lambda m_H^2 C_H + 1096\lambda^2  m_H^2  C_{H\Box} -282  \lambda^2 m_H^2  C_{HD}\right\}_2 \,,
\label{5.3}
\end{align}
\begin{align}
\dot C_{H} &=  \left\{ 108 \lambda  C_H - 160 \lambda^2  C_{H\Box} +
48\lambda^2 C_{HD}    \right\}_1  +  \left\{-3444 \lambda^2  C_H +7968  \lambda^3 C_{H\Box}  - 1992 \lambda^3  C_{HD} \right\}_2 \,,
\label{5.4}
\end{align}
\begin{align}
\dot C_{H\Box} &=  \left\{  24 \lambda C_{H\Box}\right\}_1  +  \left\{-204 \lambda^2 C_{H\Box} \right\}_2 \,,
\label{5.5}
\end{align}
\begin{align}
\dot C_{HD} &=  \left\{ 12\lambda C_{HD}    \right\}_1  +\left\{ -144 \lambda^2 C_{HD}  \right\}_2 \,,
\label{5.6}
\end{align}
\begin{align}
\dot C_{r} &=  \left\{ 12 \lambda C_{r}   \right\}_1  + \left\{ -60 \lambda^2 C_{r} \right\}_2 \,, & r & \in \{HG,H\widetilde G,HW, H\widetilde W, HB, H\widetilde B\}  \,,
\label{5.7}
\end{align}
\begin{align}
\dot C_{r} &=  \left\{ 4 \lambda C_{r}   \right\}_1  + \left\{ -28 \lambda^2 C_{r} \right\}_2 \,, &  r & \in\{HWB, H\widetilde WB\} \,.
\label{5.8}
\end{align}
The one-loop contributions agree with the known values~\cite{Jenkins:2013wua,Jenkins:2013zja,Alonso:2013hga}. The two loop contributions with the dimension-six coefficients set to zero agree with the known two loop values for the SM~\cite{Machacek:1983tz,Machacek:1983fi,Machacek:1984zw}. Note that there are very large coefficients in the two-loop anomalous dimensions. The 't~Hooft consistency conditions for the couplings listed above are satisfied.

The Higgs field anomalous dimension is divergent at two loops, i.e.\ the  't~Hooft consistency conditions are not satisfied,
\begin{align}
\gamma_H &= \left\{ 3  m_H^2 C_{H\Box} \right\}_1 + \left\{ - \frac{1}{\epsilon} 18  \lambda m_H^2 C_{H\Box}+ 6 \lambda^2 - 8 \lambda m_H^2  C_{H\Box} +2 \lambda m_H^2  C_{HD}   \right\}_2
\label{3}
\end{align}
and the divergent term agrees with eq.~\eqref{4.19} for $n=4$ on replacing $m^2$ and $C_E$ by their equivalent SMEFT couplings, $m^2 \to - m_H^2/2$, $C_E \to C_{H\Box}/2$.

\section{$\chi$PT}\label{sec:chpt}

The final example we consider is chiral perturbation theory.  We study the counterterms for the QCD chiral lagrangian for $n$ flavors, with the symmetry breaking pattern $SU(n) \times SU(n) \to SU(n)$. The notation follows ref.~\cite{Bijnens:1999sh,Bijnens:1999hw}, except that we use a Minkowski signature for the spacetime metric. We  restrict ourselves to the pure chiral theory without external sources. The chiral Lagrangian is written in terms of a field $u$,
\begin{align}
u(x) &= e^{\frac{i}{f} \bm{\pi}(x) } & \bm{\pi}(x) &= \pi^a(x)T^a & \Tr{ T^a T^b } &= \frac12 \delta^{ab}
\label{6.5}
\end{align}
where $f \sim 93\,\text{MeV}$ is the pion decay constant, and $\Tr{\, \cdot \, }$ denotes a trace. Under chiral $SU(n) \times SU(n)$, $u(x)$ transforms as
\begin{align}
u(x) &\to R u(x) h^{-1} (x) = h(x) u(x) L^{-1}
\label{6.6}
\end{align}
where $L$ and $R$ are the left- and right-handed $SU(n)$ transformations, and $h(x)$ is defined implicitly through eq.~\eqref{6.6}.
The field $U(x) = u(x)^2$ transforms as $U(x) \to R U(x) L^{-1} $. Note that $L$ and $R$ are global transformations, and do not depend on $x$, since we have not included external gauge fields.

Chiral perturbation theory has a systematic expansion in powers of $p^2$, where $p$ is the external momentum of the pions. The leading term is the order $p^2$ Lagrangian,
\begin{align}
\mathcal{L}_{2} &= \frac{f^2}{4} \vev{u_\mu\, u^\mu}\,,
\label{6.1}
\end{align}
where
\begin{align}
u_\mu = i (u^\dagger \, \partial_\mu u- u\, \partial_\mu u^\dagger)\,.
\label{6.2}
\end{align} 
The order $p^4$ Lagrangian is
\begin{align}
\mathcal{L}_{4} &=  \widehat L_0 \vev{u_\mu u_\nu u_\mu u_\nu} + \widehat L_1 \vev {u \cdot u}^2 + \widehat L_2 \vev {u_\mu u_\nu} \vev{u_\mu u_\nu} + \widehat L_3 \vev{(u \cdot u)^2}
\label{6.7}
\end{align}
with four independent coefficients in the absence of external sources. The coefficients in Eq.~\eqref{6.7} include counterterm contributions,
\begin{align}
\widehat L_i &=(c \mu)^{-2\epsilon} \left[ - \frac{1}{2\epsilon} \frac{1}{16 \pi^2} \widehat \Gamma_i + \widehat L_i^r(\mu) \right] \,,
\label{6.8}
\end{align}
where $\widehat \Gamma_i$ are the counterterms,  $\widehat L_i^r(\mu)$ are the finite renormalized couplings, and
$c^2 \mu^2 = \mu^2 e^{\gamma-1}/(4\pi)$. The renormalized coefficients satisfy the renormalization group equations
\begin{align}
\mu\, \frac{ \rd\widehat L_i^r}{\rd \mu} &=  - \frac{1}{16 \pi^2}   \widehat \Gamma_i  \,.
\label{6.21}
\end{align}

The $p^6$ Lagrangian is
\begin{align}
\mathcal{L}_{6} &=  \sum_{i=1}^{115} K_i Y_i
\label{6.3}
\end{align}
with coefficients $K_i$ and operators $Y_i$ listed in ref.~\cite{Bijnens:1999hw}. The coefficients $K_i$ include counterterm contributions, and are given in terms of renormalized couplings $K_i^r$ by
\begin{align}
K_i &= \frac{(c \mu)^{-4\epsilon}}{f^2} \left[ -\frac{1}{4 \epsilon^2}  \frac{1}{(16 \pi^2)^2}  \widehat \Gamma^{(2)}_i 
+ \frac{1}{2 \epsilon }  \frac{1}{16 \pi^2}   \widehat \Gamma^{(1)}_i + \frac{1}{2 \epsilon } \frac{1}{16 \pi^2}  \widehat \Gamma^{(L)}_i + K_i^r \right]
\label{norm}
\end{align}
$ \widehat \Gamma^{(1,2)}_i $ are the counterterms for two-loop graphs from the $p^2$ Lagrangian, and $ \widehat \Gamma^{(L)}_i$ are the counterterms from one-loop graphs with an insertion of the $p^4$ Lagrangian.
The renormalized couplings satisfy the renormalization group equations
\begin{align}
\mu\, \frac{ \rd K_i^r}{\rd \mu}  &=  \frac{1}{16 \pi^2}  \left[ 2  \widehat \Gamma^{(1)}_i  + \widehat \Gamma^{(L)}_i  \right] \,.
\label{6.22}
\end{align}
't~Hooft's consistency conditions for the $1/\epsilon^2$ pole  are
\begin{align}
 \mu \frac{\rd  \widehat \Gamma^{(L)}_i }{\rd \mu} &= -\frac{1}{8 \pi^2}  \widehat \Gamma^{(2)}_i 
 \label{6.23}
\end{align}
which become
\begin{align}
2 \, \widehat \Gamma^{(2)}_i  &= \sum_k \frac{\partial  \widehat \Gamma^{(L)}_i }{\partial \widehat L^r_k}\, \widehat \Gamma_k 
 \label{6.24}
\end{align}
using eq.~\eqref{6.21}. Eqs.~\eqref{6.23}, \eqref{6.24} are known as
Weinberg's consistency conditions~\cite{Weinberg:1978kz} for $\chi$PT, and provide a check on the two-loop result.

In the presence of external sources, there are 115 terms in the $p^6$ Lagrangian. There are 21 operators listed in ref.~\cite{Bijnens:1999hw} which are non-zero when external sources are turned off --- $Y_{1-6}$ and $Y_{49-63}$. However, in the absence of external sources, these operators are not all linearly independent. There are two linear relations,\footnote{The existence of two relations was noted in Ref.~\cite{Webber:2008aa,Bijnens:2018lez,Graf:2020yxt}. The linear relations can be obtained from Table~3.13 in~\cite{Webber:2008aa} after replacing relation 26 by $2 \times (26)+(63)+2 \times (76) + (137) = 0$.  }
\begin{align}
0 &= Y_2 + 2 Y_4  + 6 Y_6 - 2Y_{50} + 2 Y_{57} \,,  \nn
0 &= 4 Y_1 -   Y_3 + 3 Y_5  -4 Y_{49} - 2  Y_{52} + 8  Y_{54}+ 2  Y_{58}  - 4 Y_{60} \,,
\label{6.4}
\end{align}
leaving 19 independent operators. In the presence of external sources, the linear combinations in eq.~\eqref{6.4} are proportional to operators involving external sources, on using field redefinitions. The choice in Ref.~\cite{Bijnens:1999hw} was to include the linear combinations in eq.~\eqref{6.4} rather than the corresponding source-dependent operators. We have used eq.~\eqref{6.4} to eliminate $Y_{57}$ and $Y_{60}$.

The first relation in eq.~\eqref{6.4} explains one feature of the counterterms found in ref.~\cite{Bijnens:1999hw}. There is a contribution to $\widehat \Gamma^{(L)}$ proportional to $\widehat L_9^r$,
\begin{align}
\widehat \Gamma^{(L)}_i Y_i &= \widehat L_9^r \left( -\frac18 Y_2 -\frac14 Y_4 - \frac34 Y_6 + \frac14 Y_{50} - \frac14 Y_{57} \right)\,,
\label{6.20}
\end{align}
from Table~V in ref.~\cite{Bijnens:1999hw}. $\widehat L_9$ is not a coefficient in the $p^4$ Lagrangian eq.~\eqref{6.7}, and should not contribute to the counterterm. It multiplies the operator $i \Tr{ f_{+\mu \nu} u_\mu u_\nu}$ in the $p^4$ Lagrangian, which vanishes when there are no external fields. The linear combination eq.~\eqref{6.20} vanishes on using eq.~\eqref{6.4}.

Weinberg's power counting rule for $\chi$PT implies that the order $p^2$ scattering amplitude is given by tree graphs with vertices from the $p^2$ Lagrangian. The order $p^4$ scattering amplitude is given by tree graphs with one insertion of the $p^4$ Lagrangian and arbitrary $p^2$ vertices, or one-loop graphs with only $p^2$ vertices. The $p^6$ amplitude is given by tree graphs with one insertion of the $p^6$ Lagrangian or two insertions of the $p^4$ Lagrangian (and arbitrary $p^2$ vertices), one-loop graphs with one insertion of the $p^4$ Lagrangian (and arbitrary $p^2$ vertices), or two-loop graphs involving only $p^2$ interactions.

The fields in $\chi$PT live on a coset space $G/H$, which is the group manifold $SU(n)$ for QCD chiral perturbation theory. The independent coordinates can be chosen to be $\pi^a$ defined in eq.~\eqref{6.5}, and the pion covariant derivative~\cite{Coleman:1969sm,Callan:1969sn} is
\begin{align}
D_\mu \pi &= -\frac{f}{2}\, u_\mu = \partial_\mu \pi - \frac{1}{6f^2} [\pi, [\pi,\partial_\mu \pi]] + \ldots
\label{6.10}
\end{align}
The Riemann curvature tensor is
\begin{align}
R_{abcd} &= \frac{1}{f^2} f_{abg} f_{cdg} + \ldots
\label{6.11}
\end{align}
In the notation of ref.~\cite{Bijnens:1999sh,Bijnens:1999hw}, the chiral covariant derivative acting on a field $X$ that transforms as $X \to h X h^{-1}$ is
\begin{align}
\nabla_\mu X &= \partial_\mu X + \left[\Gamma_\mu,X \right] & \Gamma_\mu &= \frac12 ( u^\dagger \partial_\mu u + u \partial_\mu u^\dagger )
\label{6.12}
\end{align}
and the curvature (field-strength) is
\begin{align}
[\nabla_\mu , \nabla_\nu] X &= \left[\Gamma_{\mu \nu}, X \right] & \Gamma_{\mu \nu} = \frac14 \left[u_\mu , u_\nu \right]
-\frac{i}2 f_{+\mu \nu}
\label{6.13}
\end{align}
where $ f_{+\mu \nu} =0$ when external sources are turned off. $\Gamma_{\mu \nu}$ is the curvature eq.~\eqref{6.11} on rescaling by $(2/f)^2$ from the normalization of $D_\mu \pi$ relative to $u_\mu$.

For a Lie group, the covariant derivative of the Riemann curvature tensor vanishes, 
\begin{align}
\nabla_e R_{abcd}=0\,.
\label{6.14}
\end{align}
This greatly simplifies the computation of loop corrections, since many terms in the tensors $X,Y,A,A^\mu,B, B^\mu,B^{\mu\nu}$ and their covariant derivatives vanish. With external sources set to zero, the tensors take the simple form:
\begin{align}
X_{ab}  &= -R_{acbd} (D_\mu \pi)^c (D^\mu \pi)^d  \,, \nn
[Y_{\mu \nu}]_{ab}  &= R_{abcd} (D_\mu \pi)^c (D_\nu \pi)^d \,, \nn
A_{abc} &=  0 \,, \nn
A^\mu_{a | bc} &=  \frac13 \left( R_{abcd}  + R_{acbd}  \right) (D^\mu \pi)^d  \nn
A^{\mu \nu}_{ab | c} &= 0 \nn
B_{abcd} &=   \frac16  R_{eabf}   R_{ecdg} (D_\mu \pi)^f (D^\mu \pi)^g \quad  \text{sym}(abcd) \,, \nn
B^\mu _{a | bcd} &=0  \nn
B^{\mu \nu}_{ab | cd} &=  -\frac1{12}  \eta^{\mu \nu} \left( R_{acbd} + R_{adbc} \right) \,.
\label{6.16}
\end{align}

The one-loop graphs from the $p^2$ Lagrangian generate order $p^4$ counterterms using eq.~\eqref{3.5}. We can expand about $\pi=0$, and compare our counterterms with eq.~\eqref{6.7} to get the one-loop counterterm coefficients
\begin{align}
\widehat \Gamma_0 &= \frac{n}{48}\,, &
\widehat \Gamma_1 &= \frac{1}{16} \,, &
\widehat \Gamma_2 &= \frac{1}{8} \,, &
\widehat \Gamma_3 &= \frac{n}{24} \,,
\label{6.15}
\end{align}
which agrees with \cite[(3.14)]{Bijnens:1999hw}.

The two-loop graphs from the $p^2$ Lagrangian generate order $p^6$ counterterms $\widehat \Gamma^{(2)}_i$ and $\widehat \Gamma^{(1)}_i$ which are listed in Table~\ref{tab:chpt}. These agree with Table~IV of \cite{Bijnens:1999hw} using the identities eq.~\eqref{6.4} to eliminate $Y_{57}$ and $Y_{60}$. The agreement provides a highly non-trivial check of our method.
%
%
\begin{table}
\renewcommand{\arraycolsep}{0.5cm}
\renewcommand{\arraystretch}{1.5}
\begin{align}
\begin{array}{c|c|c|cc}
Y &  \widehat \Gamma^{(2)}_i & 16 \pi^2 \widehat \Gamma^{(1)}_i  & \widehat\Gamma^{(L)}_i  \nn
\hline
1 &  \frac{5}{48} + \frac{1}{64} n^2 &- \frac{67}{576} - \frac{7}{1728} n^2 &  \frac{1}{12} n  \widehat L_0 +3  \widehat L_1 + \frac16 \widehat L_2  + \frac{17}{24} n \widehat L_3 \\
2 &  \frac{1}{576} n & -\frac{31}{6912} n &  \frac{5}{24} \widehat L^r_0 - \frac1{24} n \widehat L^r_2  + \frac{5}{48}  \widehat L^r_3 \\
3 &  - \frac{5}{48} + \frac{1}{2304} n^2 & \frac{61}{2304} + \frac{11}{27648} n^2 &  \frac{1}{48} n  \widehat L_0 - \frac{7}{6} \widehat L_1 - \frac{13}{12} \widehat L_2  + \frac{1}{96} n \widehat L_3   \\
4 & - \frac{11}{72} n & - \frac{49}{3456} n &  -\frac{5}{4}   \widehat L^r_0 -2 n \widehat L^r_1 - \frac34 n \widehat L^r_2  - \frac{35}{24} \widehat L^r_3  \\
5 &   -\frac{1}{768}n^2 & -\frac{23}{256} - \frac{49}{27648} n^2 &  - \frac{11}{48} n  \widehat L_0 + \frac{11}{6} \widehat L_1 - \frac{11}{12} \widehat L_2  + \frac{5}{96} n \widehat L_3 \\
6 & -\frac{13}{32} n & -\frac{5}{192} n & - \frac{3}{4}  \widehat L^r_0 - 6 n  \widehat L^r_1 - \frac94 n \widehat L^r_2  - \frac{27}{8}  \widehat L^r_3  \\
49 & - \frac{5}{576} n^2 & \frac{5}{48} + \frac{1}{2304} n^2  &  \frac{1}{4} n  \widehat L_0 - \frac{4}{3} \widehat L_1 + \frac23 \widehat L_2  - \frac{13}{24} n \widehat L_3  \\
50 &  \frac{1}{32} n &  \frac{5}{128} n &  -\frac{1}{4}   \widehat L^r_0 + \frac{1}{4} n \widehat L^r_2  + \frac{7}{8}  \widehat L^r_3  \\
51 &   \frac{1}{64} & \frac{5}{256} &  \frac14 \widehat L^r_2 \\
52 &  - \frac{11}{384} n^2 &  \frac{17}{128} +\frac{77}{13824} n^2 &    -\frac{17}{24} n  \widehat L_0 - \frac{1}{3} \widehat L_1 + \frac16 \widehat L_2  - \frac{49}{48} n \widehat L_3 \\
53 &  - \frac{1}{64} n & - \frac{5}{256} n & \widehat L^r_0 - \frac{5}{4} \widehat L^r_3 \\
54  & \frac{1}{48} n^2 &  -\frac{17}{64} - \frac{13}{3456} n^2 &   -\frac{1}{3} n  \widehat L_0 + \frac{2}{3} \widehat L_1 - \frac13 \widehat L_2  + \frac{7}{6} n \widehat L_3  \\
55 &  - \frac{1}{24} n & - \frac{1}{72} n & - \frac{2}{3}   \widehat L^r_0  - \frac{2}{3} n \widehat L^r_2  + \frac{1}{3}  \widehat L^r_3  \\
56 &  - \frac{1}{32}  &  \frac{3}{128} & -\frac12 \widehat L^r_2 \\
58 &    \frac{1}{1152} n^2 &  \frac{11}{384}-  \frac{13}{13824} n^2 &   \frac{1}{24} n  \widehat L_0 +  \widehat L_1 - \frac12 \widehat L_2  + \frac{1}{48} n \widehat L_3 \\
59 &  - \frac{1}{192} n &  \frac{65}{2304} n & \widehat L^r_0 - \frac34 \widehat L^r_3 \\
61 &   \frac{7}{192} n & - \frac{23}{2304} n  &  \widehat L^r_0 + \frac54 \widehat L^r_3  \\
62 &   - \frac{1}{12} n &  - \frac{5}{288} n & -\frac{13}{3} \widehat L^r_0 -\frac13 n \widehat L^r_2 - \frac56 \widehat L^r_3 \\
63 &   - \frac{1}{8} &  -\frac{1}{32} & -2 \widehat L^r_2 \\
\end{array}
\end{align}
\caption{\label{tab:chpt} The counterterms for the $p^6$ Lagrangian. The first column lists the operator $Y_i$. $Y_{57}$ and $Y_{60}$ were eliminated using eq.~\eqref{6.4}. The second and third columns list the $1/\epsilon^2$ and $1/\epsilon$  counterterm coefficients $\widehat \Gamma^{(2)}_i$ and  $\widehat \Gamma^{(1)}_i$ from two-loop graphs from the $p^2$ Lagrangian. The last column lists the $1/\epsilon$ one-loop counterterm from an insertion of the $p^4$ Lagrangian.  The normalization is given in Eq.~\eqref{norm}. The counterterms agree with ref.~\cite{Bijnens:1999hw} after using eq.~\eqref{6.4}. }
\end{table}
%
%

We also need the one-loop corrections with an insertion of the $p^4$ Lagrangian eq.~\eqref{6.7}. The $p^4$ Lagrangian has four derivatives, but at most one derivative acting on a single field, and our method can still be used. The one-loop graphs are computed by expanding the Lagrangian to second order in the quantum field $\eta$, i.e.\ by taking the second derivative of the Lagrangian in the Riemann normal coordinate expansion of Sec.~\ref{sec:riemann}. Using the basic equations eq.~\eqref{2.13} and eq.~\eqref{2.18} for the derivatives on fields, the second variation of the $p^4$ Lagrangian is

\begin{align}
C_{ab} \eta^a \eta^b &=2  \widehat  L_0   \Tr{R(\eta,u_\mu ) \eta \, u_\nu\,   u_\mu  \, u_\nu}  +2  \widehat   L_1 \Tr{R(\eta,u_\mu ) \eta \, u_\mu } \Tr{   u_\nu  \, u_\nu}  + 2  \widehat   L_2  \Tr{R(\eta,u_\mu ) \eta \, u_\nu } \Tr{   u_\mu  \, u_\nu}\nn
& + \widehat   L_3 \bigl[   \Tr{R(\eta,u_\mu ) \eta \, u_\mu\,   u_\nu  \, u_\nu} +  \Tr{R(\eta,u_\mu ) \eta \, u_\nu\,   u_\nu  \, u_\mu}  \bigr]
\end{align}
\begin{align}
C^\mu_{ab} &= 0
\end{align}
\begin{align}
C^{\mu \nu}_{ab} & (\nabla_\mu \eta)^a(\nabla_\nu \eta)^b
 = \widehat  L_0 \left[ 4 \Tr{\nabla_\mu \eta\, \nabla_\nu \eta \, u_\mu \,  u_\nu} + 2 \Tr{\nabla_\mu \eta\,  u_\nu\, \nabla_\mu  \eta \, u_\nu} 
  \right] \nn
& + \widehat   L_1 \left[ 2 \Tr{\nabla_\mu \eta\, \nabla_\mu \eta } \Tr{ u_\nu \,  u_\nu} + 4 \Tr{\nabla_\mu \eta\, u_\mu  } \Tr{ \nabla_\nu \eta\,  u_\nu}
  \right] \nn
& + \widehat   L_2 \left[  2\Tr{\nabla_\mu \eta\, \nabla_\nu \eta } \Tr{ u_\mu \,  u_\nu} + 2 \Tr{\nabla_\mu \eta\, u_\nu  } \Tr{ \nabla_\mu \eta\,  u_\nu} + 2  \Tr{\nabla_\mu \eta\, u_\nu  } \Tr{ \nabla_\nu \eta\,  u_\mu}
 \right] \nn
& + \widehat   L_3 \bigl[  2\Tr{\nabla_\mu \eta\, \nabla_\mu \eta \, u_\nu \,  u_\nu} + 2 \Tr{\nabla_\mu \eta\, \nabla_\nu \eta \, u_\nu \,  u_\mu}   +  \Tr{\nabla_\mu \eta\,u_\mu \,   \nabla_\nu \eta \,  u_\nu}  +  \Tr{\nabla_\mu \eta\,u_\nu \,   \nabla_\nu \eta \,  u_\mu}   \bigr] 
\label{6.18}
\end{align}
on comparing with the general second variation eq.~(3.10) in paper I. The one-loop counterterm in eq.~(3.11) of paper I gives the results in the $\widehat \Gamma^{(L)}_i$ column of Table~\ref{tab:chpt}, and agrees with \cite{Bijnens:1999hw} using eq.~\eqref{6.4}.\footnote{In an earlier version of this paper, we found a discrepancy with the result of \cite{Bijnens:1999hw}. The source of this discrepancy was that in eq.~(3.11) of paper I, the terms $\frac1{12}  C^{\mu \nu}_{ab} Y^{\mu \alpha}_{bc} Y^{\nu \alpha}_{ca} 
- \frac1{4}  C^{\mu \nu}_{ab} Y^{\nu \alpha}_{bc} Y^{\mu \alpha}_{ca}$ were instead written as $\frac1{12}  C^{\mu \nu}_{ab} Y^{\nu \alpha}_{bc} Y^{\mu \alpha}_{ca} 
- \frac1{4}  C^{\mu \nu}_{ab} Y^{\mu \alpha}_{bc} Y^{\nu \alpha}_{ca}$. We thank J.~Bijnens, G.~Colangelo, and G.~Ecker for help resolving this discrepancy.}

In the general case, one-loop corrections from higher derivative terms with at most a single derivative on each field can be computed the same way. First, the fluctuation Lagrangian is computed to quadratic order in $\eta$ to determine the coefficients $C_{ab}$, $C^\mu_{ab}$ and $C^{\mu \nu}_{ab}$, and then, the one-loop counterterms are computed from eq.~(3.11) of paper I.  This was also used to compute the one-loop RGE in SMEFT from dimension-eight $H^4D^4$ operators, which agrees with the previous calculation of ref.~\cite{DasBakshi:2022mwk} using diagrammatic methods. The $H^4D^4$ contribution to the  SMEFT RGEs is given in Appendix~\ref{app:h4d4}.

Two-loop corrections are computed from $\eta^3$ and $\eta^4$ fluctuation terms. Higher derivative interactions can generate $(D\eta)^2 \eta$, $(D\eta)^3$, $(D\eta)^3\eta$ and $(D\eta)^4$ interactions. The two-loop counterterms can then be computed by the same method as given in paper I for the $A$-type and $B$-type two-loop counterterms. However, we have not explicitly computed the counterterm coefficients with these additional interactions, so we cannot determine the two-loop corrections from higher derivative operators without additional work.

\section{Conclusions}\label{sec:conc}

The geometric method has been used to compute the two-loop renormalization of generic scalar QFTs, applying the counterterm formula developed in paper I. The counterterms can be derived purely algebraically, in terms of covariant derivatives of the scalar potential and Riemann curvature tensor, which are computed by taking derivatives of polynomial coefficients in the Lagrangian. We have applied the method to the $O(n)$ EFT, the Higgs sector of the SMEFT, and chiral perturbation theory. The agreement with previous results computed by other methods provides a highly non-trivial check on our calculation.

The results in this paper are for scalar loops. The method is generalizable to gauge and fermion loops, as has already been done for one-loop renormalization~\cite{Helset:2022pde,Helset:2022tlf,Finn:2020nvn,Gattus:2023gep,Assi:2023zid}.

\subsection*{Acknowledgments}

We thank Xiaochuan Lu, Chia-Hsien Shen,  Peter Stoffer and Anders Thomsen for helpful discussions. We would also like to thank J.~Bijnens, G.~Colangelo and G.~Ecker for helpful comments about their results in ref.~\cite{Bijnens:1999hw}.
This work is supported in part by the U.S.\ Department of Energy (DOE) under award numbers~DE-SC0009919. LN gratefully acknowledges financial support by the Swiss National Science Foundation (Project No.~PCEFP2\_194272 and mobility grant PCEFP2\_194272/2).

\begin{appendix}

\section{Two-loop Counterterms for the $O(n)$ EFT}\label{app:on}

Using our one-loop and two-loop formul\ae\ gives the counterterms for the $O(N)$ EFT listed below. The subscripts $1,2$ denote the one-loop and two-loop contributions, respectively  (see footnote~\ref{foot}).
\begin{align}
Z_\phi &= 1 + \frac{1}{\epsilon}  \left\{ 4 (n-1)  m^2 C_E \right\}_1 +  \frac{1}{\epsilon^2} \left\{ 4 (n-1)(n+2) \lambda m^2 C_E \right\}_2 +
 \frac{1}{\epsilon} \left\{ -\frac12 (n+2)\lambda^2 + \frac43 (n+2) \lambda m^2 C_E \right\}_2
\label{A.2}
\end{align}
\begin{align}
\Lambda_{\text{c.t.}} &= \frac{1}{\epsilon}  \left\{ \frac{ n m^4}{4 } \right\}_1 +\frac{1}{\epsilon^2}  \left\{ \frac14 n(n+2) \lambda  m^4 - n^2  m^6 C_E   \right\}_2 
\label{A.1}
\end{align}
\begin{align}
m^2_{\text{c.t.}} &= \frac{1}{\epsilon} \left\{(n+2) \lambda  m^2 - 4 n  m^4 C_E \right\}_1 + \frac{1}{\epsilon^2} \bigl\{ 
(n+2)(n+5) \lambda^2 m^2 - 2 (n+2)(7n+6) \lambda m^4 C_E \nn
& - 6 (n+2)(n+4) m^4  C_1 \bigr\}_2 + \frac{1}{\epsilon} \left\{ -\frac52 (n+2) \lambda^2 m^2 + \frac{20}{3}(n+2) \lambda   m^4 C_E\right\}_2
\label{A.3}
\end{align}
\begin{align}
\lambda_{\text{c.t.}}  &= \frac{1}{\epsilon} \left\{(n+8) \lambda^2 - 8(n+3) \lambda m^2  C_E  - 12 (n+4) m^2 C_1 \right\}_1 + \frac{1}{\epsilon^2} \bigl\{  (n+8)^2 \lambda^3 \nn
&  - 12 (2n^2+21n+50)  \lambda^2 m^2 C_E -36 (n+4)(n+10) \lambda m^2 C_1  \bigr\}_2  \nn
& + \frac{1}{\epsilon} \left\{ -3 (3n+14) \lambda^3 + \frac83 (22n+113) \lambda^2 m^2 C_E +120 (n+4) \lambda m^2 C_1 \right\}_2
\label{A.4}
\end{align}
\begin{align}
C_{1, \text{c.t.}} &=  \frac{1}{\epsilon} \left\{  10 \lambda^2 C_E   + 3 (n+14) \lambda C_1 \right\}_1 + \frac{1}{\epsilon^2} \bigl\{ 5(7n+62) \lambda^3 C_E + 3 (n+14)(2n+25) \lambda^2 C_1   \bigr\}_2  \nn
& + \frac{1}{\epsilon} \left\{- \frac23 (23n+259) \lambda^3 C_E  - \frac{21}{2} (7n+54) \lambda^2 C_1 \right\}_2
\label{A.5}
\end{align}
\begin{align}
C_{E ,\text{c.t.}} &=  \frac{1}{\epsilon} \left\{ 2 (n+2) \lambda C_E  \right\}_1 + \frac{1}{\epsilon^2} \bigl\{ 3 (n+2)(n+4) \lambda^2 C_E  \bigr\}_2 + \frac{1}{\epsilon} \left\{ -\frac{17}{2} (n+2) \lambda^2 C_E  \right\}_2
\label{A.6}
\end{align}

\section{Two-loop Counterterms for SMEFT}\label{app:smeft}

The one- and two-loop counterterms for the scalar sector of SMEFT are
\begin{align}
Z_H &=  1 + \frac{1}{\epsilon} \left\{ - 3 C_{H\Box} m_H^2 \right\}_1 + \frac{1}{\epsilon} \left\{ -3 \lambda^2 + 4 C_{H\Box}  \lambda m_H^2
- C_{HD} \lambda m_H^2  \right\}_2
+ \frac{1}{\epsilon^2} \left\{ - 18 C_{H\Box} \lambda m_H^2 \right\}_2
\end{align}

\begin{align}
\Lambda_{\text{c.t.}} &= \frac{1}{\epsilon} \left \{ \frac14 m_H^4 \right\}_1 + \frac1{\epsilon^2} \left\{\frac32 \lambda m_H^4 -\frac12 C_{H\Box} m_H^6 + \frac14 C_{HD} m_H^6 \right\}_2
\end{align}

\begin{align}
(m_H^2)_{\text{c.t.}} &= \frac{1}{\epsilon} \left\{ 6 \lambda m_H^2- 2 C_{H\Box} m_H^4 + C_{HD} m_H^4 \right\}_1 \nn
& + \frac1{\epsilon^2} \left\{ 54 \lambda^2 m_H^2 + 18 C_H m_H^4 -78 C_{H\Box} \lambda m_H^4 + 30 C_{HD}   \lambda m_H^4\right\}_2  \nn
& + \frac1{\epsilon} \left\{ -15 \lambda^2 m_H^2 + 20  C_{H\Box} \lambda m_H^4 -5 C_{HD}   \lambda m_H^4\right\}_2 
\end{align}

\begin{align}
\lambda_{\text{c.t.}} &= \frac{1}{\epsilon} \left \{ 12 \lambda^2 + 6 C_H m_H^2 -16 C_{H\Box} \lambda m_H^2 + 6 C_{HD} \lambda m_H^2 \right\}_1 \nn
& + \frac1{\epsilon^2} \left\{ 144 \lambda^3  + 252 C_H \lambda m_H^2 - 672  C_{H\Box} \lambda^2 m_H^2 + 216  C_{HD}   \lambda^2 m_H^2\right\}_2 \nn
& + \frac1{\epsilon} \left\{ -78 \lambda^3 -60  C_H \lambda m_H^2 + 274  C_{H\Box} \lambda^2 m_H^2 - \frac{141}{2} C_{HD}   \lambda^2 m_H^2\right\}_2 
\end{align}

\begin{align}
(C_H)_{\text{c.t.}} &= \frac{1}{\epsilon} \left \{ 54 C_H \lambda - 80 C_{H\Box} \lambda^2 +24 C_{HD} \lambda^2  \right\}_1 \nn
& + \frac1{\epsilon^2} \left\{  1782 C_H \lambda^2 - 3600  C_{H\Box} \lambda^3 + 1008  C_{HD}   \lambda^3\right\}_2 \nn
& + \frac1{\epsilon} \left\{ -861  C_H \lambda^2 + 1992  C_{H\Box} \lambda^3 -498  C_{HD}   \lambda^3 \right\}_2 
\end{align}

\begin{align}
(C_{H\Box})_{\text{c.t.}} &= \frac{1}{\epsilon} \left \{ 12 C_{H\Box} \lambda  \right\}_1 + \frac1{\epsilon^2} \left\{   144 C_{H\Box} \lambda^2 \right\}_2  + \frac1{\epsilon} \left\{ -51 C_{H\Box} \lambda^2 \right\}_2  
\end{align}

\begin{align}
(C_{HD})_{\text{c.t.}} &= \frac{1}{\epsilon} \left \{ 6 C_{HD} \lambda  \right\}_1 + \frac1{\epsilon^2} \left\{   54 C_{HD} \lambda^2 \right\}_2  + \frac1{\epsilon} \left\{ -36 C_{HD} \lambda^2 \right\}_2  
\end{align}

\begin{align}
(C_{r})_{\text{c.t.}} &= \frac{1}{\epsilon} \left \{ 6 C_{HG} \lambda  \right\}_1  + \frac1{\epsilon^2} \left\{   54 C_{HG} \lambda^2 \right\}_2 + \frac1{\epsilon} \left\{ -15 C_{HG} \lambda^2 \right\}_2   
\end{align}
for $r=HG, H \widetilde G, HW, H \widetilde W, HB, H \widetilde B$.
and
\begin{align}
(C_{r})_{\text{c.t.}} &= \frac{1}{\epsilon} \left \{ 2 C_{HWB} \lambda  \right\}_1 + \frac1{\epsilon^2} \left\{   14 C_{HWB} \lambda^2 \right\}_2  + \frac1{\epsilon} \left\{ -7 C_{HWB} \lambda^2 \right\}_2  
\end{align}
also  $r=HWB, H \widetilde WB$.

\section{SMEFT $H^4 D^4$ Insertions} \label{app:h4d4}

The contributions of $H^4D^4$ insertions to the SMEFT renormalization group equations up to dimension eight (with $1/(16 \pi^2)$ absorbed into the definition of $\rd/\rd t$) are:
\begin{align}
\dot m_H^2  &= -\frac{3}{2}  m_H^6 \ \coef{8}{H^4D^4}{(1)} -\frac{3}{4}  m_H^6\ \coef{8}{H^4D^4}{(2)}  - \frac{9}{4}\ \coef{8}{H^4D^4}{(3)} \,,
\end{align}
\begin{align}
\dot \lambda  &= -\frac{19}{3} \lambda m_H^4 \ \coef{8}{H^4D^4}{(1)} -\frac{19}{6} \lambda m_H^4 \ \coef{8}{H^4D^4}{(2)}  - \frac{23}{2} \lambda m_H^4  \ \coef{8}{H^4D^4}{(3)} \,,
\end{align}
\begin{align}
\dot C_{H}  &= -18  \lambda^2 m_H^2 \ \coef{8}{H^4D^4}{(1)} -22  \lambda^2 m_H^2\ \coef{8}{H^4D^4}{(2)}  - 32 \lambda^2 m_H^2\ \coef{8}{H^4D^4}{(3)} \,,
\end{align}
\begin{align}
\dot C_{H\Box}  &= -2 \lambda m_H^2 \ \coef{8}{H^4D^4}{(1)} + 2 \lambda m_H^2\ \coef{8}{H^4D^4}{(2)}  - 6  \lambda m_H^2\ \coef{8}{H^4D^4}{(3)} \,,
\end{align}
\begin{align}
\dot C_{HD}  &=-4  \lambda m_H^2 \ \coef{8}{H^4D^4}{(1)} + 4 \lambda m_H^2\ \coef{8}{H^4D^4}{(2)} \,,
\end{align}
\begin{align}
\dcoef{8}{H^8}{} &= \frac{56}{3} \lambda^3\ \coef{8}{H^4D^4}{(1)} + \frac{184}{3} \lambda^3\ \coef{8}{H^4D^4}{(2)}  + 16 \lambda^3\ \coef{8}{H^4D^4}{(3)} \,,
\end{align}
\begin{align}
\dcoef{8}{H^6D^2}{(1)} &= \frac{64}{3} \lambda^2\ \coef{8}{H^4D^4}{(1)} -\frac{124}{3} \lambda^2\ \coef{8}{H^4D^4}{(2)}  + 44  \lambda^2\ \coef{8}{H^4D^4}{(3)} \,,
\end{align}
\begin{align}
\dcoef{8}{H^6D^2}{(2)} &=20 \lambda^2\ \coef{8}{H^4D^4}{(1)} - 20 \lambda^2\ \coef{8}{H^4D^4}{(2)} \,,
\end{align}
\begin{align}
\dcoef{8}{H^4D^4}{(1)} &= 8 \lambda\ \coef{8}{H^4D^4}{(1)} + \frac83 \lambda\ \coef{8}{H^4D^4}{(2)} + \frac83 \lambda\ \coef{8}{H^4D^4}{(3)} \,,
\end{align}
\begin{align}
\dcoef{8}{H^4D^4}{(2)} &= \frac83 \lambda\ \coef{8}{H^4D^4}{(1)} + 8 \lambda\ \coef{8}{H^4D^4}{(2)} + \frac83 \lambda\ \coef{8}{H^4D^4}{(3)} \,,
\end{align}
\begin{align}
\dcoef{8}{H^4D^4}{(3)} &= 16 \lambda\ \coef{8}{H^4D^4}{(1)} + \frac{32}{3} \lambda\ \coef{8}{H^4D^4}{(2)} + \frac{80}3 \lambda\ \coef{8}{H^4D^4}{(3)} \,.
\end{align}
These agree with ref.~\cite{DasBakshi:2022mwk}.

\end{appendix}

\bibliographystyle{JHEP}
\bibliography{refs.bib}

\providecommand{\href}[2]{#2}\begingroup\raggedright\begin{thebibliography}{10}

\bibitem{Alonso:2016oah}
R.~Alonso, E.~E. Jenkins and A.~V. Manohar, \emph{{Geometry of the Scalar
  Sector}}, \href{https://doi.org/10.1007/JHEP08(2016)101}{\emph{JHEP}
  {\bfseries 08} (2016) 101}
  [\href{https://arxiv.org/abs/1605.03602}{{\ttfamily 1605.03602}}].

\bibitem{Alonso:2015fsp}
R.~Alonso, E.~E. Jenkins and A.~V. Manohar, \emph{{A Geometric Formulation of
  Higgs Effective Field Theory: Measuring the Curvature of Scalar Field
  Space}}, \href{https://doi.org/10.1016/j.physletb.2016.01.041}{\emph{Phys.
  Lett. B} {\bfseries 754} (2016) 335}
  [\href{https://arxiv.org/abs/1511.00724}{{\ttfamily 1511.00724}}].

\bibitem{Helset:2022pde}
A.~Helset, E.~E. Jenkins and A.~V. Manohar, \emph{{Renormalization of the
  Standard Model Effective Field Theory from geometry}},
  \href{https://doi.org/10.1007/JHEP02(2023)063}{\emph{JHEP} {\bfseries 02}
  (2023) 063} [\href{https://arxiv.org/abs/2212.03253}{{\ttfamily
  2212.03253}}].

\bibitem{Helset:2022tlf}
A.~Helset, E.~E. Jenkins and A.~V. Manohar, \emph{{Geometry in scattering
  amplitudes}}, \href{https://doi.org/10.1103/PhysRevD.106.116018}{\emph{Phys.
  Rev. D} {\bfseries 106} (2022) 116018}
  [\href{https://arxiv.org/abs/2210.08000}{{\ttfamily 2210.08000}}].

\bibitem{Finn:2020nvn}
K.~Finn, S.~Karamitsos and A.~Pilaftsis, \emph{{Frame covariant formalism for
  fermionic theories}},
  \href{https://doi.org/10.1140/epjc/s10052-021-09360-w}{\emph{Eur. Phys. J. C}
  {\bfseries 81} (2021) 572}
  [\href{https://arxiv.org/abs/2006.05831}{{\ttfamily 2006.05831}}].

\bibitem{Gattus:2023gep}
V.~Gattus and A.~Pilaftsis, \emph{{Minimal Supergeometric Quantum Field
  Theories}},  \href{https://arxiv.org/abs/2307.01126}{{\ttfamily 2307.01126}}.

\bibitem{Assi:2023zid}
B.~Assi, A.~Helset, A.~V. Manohar, J.~Pag\`es and C.-H. Shen, \emph{{Fermion
  Geometry and the Renormalization of the Standard Model Effective Field
  Theory}},  \href{https://arxiv.org/abs/2307.03187}{{\ttfamily 2307.03187}}.

\bibitem{Polyakov:2010pt}
M.~V. Polyakov and A.~A. Vladimirov, \emph{{Leading Infrared Logarithms for
  Sigma-Model with Fields on Arbitrary Riemann Manifold}},
  \href{https://doi.org/10.1007/s11232-011-0126-7}{\emph{Theor. Math. Phys.}
  {\bfseries 169} (2011) 1499}
  [\href{https://arxiv.org/abs/1012.4205}{{\ttfamily 1012.4205}}].

\bibitem{Buchalla:2019wsc}
G.~Buchalla, A.~Celis, C.~Krause and J.-N. Toelstede, \emph{{Master Formula for
  One-Loop Renormalization of Bosonic SMEFT Operators}},
  \href{https://arxiv.org/abs/1904.07840}{{\ttfamily 1904.07840}}.

\bibitem{Helset:2020yio}
A.~Helset, A.~Martin and M.~Trott, \emph{{The Geometric Standard Model
  Effective Field Theory}},
  \href{https://doi.org/10.1007/JHEP03(2020)163}{\emph{JHEP} {\bfseries 03}
  (2020) 163} [\href{https://arxiv.org/abs/2001.01453}{{\ttfamily
  2001.01453}}].

\bibitem{Cheung:2022vnd}
C.~Cheung, A.~Helset and J.~Parra-Martinez, \emph{{Geometry-kinematics
  duality}}, \href{https://doi.org/10.1103/PhysRevD.106.045016}{\emph{Phys.
  Rev. D} {\bfseries 106} (2022) 045016}
  [\href{https://arxiv.org/abs/2202.06972}{{\ttfamily 2202.06972}}].

\bibitem{Cohen:2022uuw}
T.~Cohen, N.~Craig, X.~Lu and D.~Sutherland, \emph{{On-Shell Covariance of
  Quantum Field Theory Amplitudes}},
  \href{https://doi.org/10.1103/PhysRevLett.130.041603}{\emph{Phys. Rev. Lett.}
  {\bfseries 130} (2023) 041603}
  [\href{https://arxiv.org/abs/2202.06965}{{\ttfamily 2202.06965}}].

\bibitem{Craig:2023hhp}
N.~Craig, Yu-Tse and Lee, \emph{{Effective Field Theories on the Jet Bundle}},
  \href{https://arxiv.org/abs/2307.15742}{{\ttfamily 2307.15742}}.

\bibitem{Alminawi:2023qtf}
M.~Alminawi, I.~Brivio and J.~Davighi, \emph{{Jet Bundle Geometry of Scalar
  Field Theories}},  \href{https://arxiv.org/abs/2308.00017}{{\ttfamily
  2308.00017}}.

\bibitem{Alonso:2021rac}
R.~Alonso and M.~West, \emph{{Roads to the Standard Model}},
  \href{https://doi.org/10.1103/PhysRevD.105.096028}{\emph{Phys. Rev. D}
  {\bfseries 105} (2022) 096028}
  [\href{https://arxiv.org/abs/2109.13290}{{\ttfamily 2109.13290}}].

\bibitem{Alonso:2022ffe}
R.~Alonso and M.~West, \emph{{On the effective action for scalars in a general
  manifold to any loop order}},
  \href{https://arxiv.org/abs/2207.02050}{{\ttfamily 2207.02050}}.

\bibitem{tHooft:1973bhk}
G.~'t~Hooft, \emph{{An algorithm for the poles at dimension four in the
  dimensional regularization procedure}},
  \href{https://doi.org/10.1016/0550-3213(73)90263-0}{\emph{Nucl. Phys. B}
  {\bfseries 62} (1973) 444}.

\bibitem{Jenkins:2023rtg}
E.~E. Jenkins, A.~V. Manohar, L.~Naterop and J.~Pag\`es, \emph{{An Algebraic
  Formula for Two Loop Renormalization of Scalar Quantum Field Theory}},
  \href{https://arxiv.org/abs/2308.06315}{{\ttfamily 2308.06315}}.

\bibitem{Cao:2021cdt}
W.~Cao, F.~Herzog, T.~Melia and J.~R. Nepveu, \emph{{Renormalization and
  non-renormalization of scalar EFTs at higher orders}},
  \href{https://doi.org/10.1007/JHEP09(2021)014}{\emph{JHEP} {\bfseries 09}
  (2021) 014} [\href{https://arxiv.org/abs/2105.12742}{{\ttfamily
  2105.12742}}].

\bibitem{Cao:2023adc}
W.~Cao, F.~Herzog, T.~Melia and J.~Roosmale~Nepveu, \emph{{Non-linear
  non-renormalization theorems}},
  \href{https://doi.org/10.1007/JHEP08(2023)080}{\emph{JHEP} {\bfseries 08}
  (2023) 080} [\href{https://arxiv.org/abs/2303.07391}{{\ttfamily
  2303.07391}}].

\bibitem{Machacek:1984zw}
M.~E. Machacek and M.~T. Vaughn, \emph{{Two Loop Renormalization Group
  Equations in a General Quantum Field Theory. 3. Scalar Quartic Couplings}},
  \href{https://doi.org/10.1016/0550-3213(85)90040-9}{\emph{Nucl. Phys. B}
  {\bfseries 249} (1985) 70}.

\bibitem{Machacek:1983fi}
M.~E. Machacek and M.~T. Vaughn, \emph{{Two Loop Renormalization Group
  Equations in a General Quantum Field Theory. 2. Yukawa Couplings}},
  \href{https://doi.org/10.1016/0550-3213(84)90533-9}{\emph{Nucl. Phys. B}
  {\bfseries 236} (1984) 221}.

\bibitem{Machacek:1983tz}
M.~E. Machacek and M.~T. Vaughn, \emph{{Two Loop Renormalization Group
  Equations in a General Quantum Field Theory. 1. Wave Function
  Renormalization}},
  \href{https://doi.org/10.1016/0550-3213(83)90610-7}{\emph{Nucl. Phys. B}
  {\bfseries 222} (1983) 83}.

\bibitem{Bijnens:1999hw}
J.~Bijnens, G.~Colangelo and G.~Ecker, \emph{{Renormalization of chiral
  perturbation theory to order $p^6$}},
  \href{https://doi.org/10.1006/aphy.1999.5982}{\emph{Annals Phys.} {\bfseries
  280} (2000) 100} [\href{https://arxiv.org/abs/hep-ph/9907333}{{\ttfamily
  hep-ph/9907333}}].

\bibitem{Honerkamp:1971sh}
J.~Honerkamp, \emph{{Chiral multiloops}},
  \href{https://doi.org/10.1016/0550-3213(72)90299-4}{\emph{Nucl. Phys. B}
  {\bfseries 36} (1972) 130}.

\bibitem{Honerkamp:1971xtx}
J.~Honerkamp and K.~Meetz, \emph{{Chiral-invariant perturbation theory}},
  \href{https://doi.org/10.1103/PhysRevD.3.1996}{\emph{Phys. Rev. D} {\bfseries
  3} (1971) 1996}.

\bibitem{Eisenhart:1949vn}
L.~P. Eisenhart, \emph{Riemannian Geometry}. Princeton University Press, 1949.

\bibitem{Bednyakov:2014pia}
A.~V. Bednyakov, A.~F. Pikelner and V.~N. Velizhanin, \emph{{Three-loop SM
  beta-functions for matrix Yukawa couplings}},
  \href{https://doi.org/10.1016/j.physletb.2014.08.049}{\emph{Phys. Lett. B}
  {\bfseries 737} (2014) 129}
  [\href{https://arxiv.org/abs/1406.7171}{{\ttfamily 1406.7171}}].

\bibitem{Herren:2017uxn}
F.~Herren, L.~Mihaila and M.~Steinhauser, \emph{{Gauge and Yukawa coupling beta
  functions of two-Higgs-doublet models to three-loop order}},
  \href{https://doi.org/10.1103/PhysRevD.97.015016}{\emph{Phys. Rev. D}
  {\bfseries 97} (2018) 015016}
  [\href{https://arxiv.org/abs/1712.06614}{{\ttfamily 1712.06614}}], [Erratum:
  Phys.Rev.D 101, 079903 (2020)].

\bibitem{Herren:2021yur}
F.~Herren and A.~E. Thomsen, \emph{{On ambiguities and divergences in
  perturbative renormalization group functions}},
  \href{https://doi.org/10.1007/JHEP06(2021)116}{\emph{JHEP} {\bfseries 06}
  (2021) 116} [\href{https://arxiv.org/abs/2104.07037}{{\ttfamily
  2104.07037}}].

\bibitem{Jenkins:2013wua}
E.~E. Jenkins, A.~V. Manohar and M.~Trott, \emph{{Renormalization Group
  Evolution of the Standard Model Dimension Six Operators II: Yukawa
  Dependence}}, \href{https://doi.org/10.1007/JHEP01(2014)035}{\emph{JHEP}
  {\bfseries 01} (2014) 035} [\href{https://arxiv.org/abs/1310.4838}{{\ttfamily
  1310.4838}}].

\bibitem{Jenkins:2013zja}
E.~E. Jenkins, A.~V. Manohar and M.~Trott, \emph{{Renormalization Group
  Evolution of the Standard Model Dimension Six Operators I: Formalism and
  $\lambda$ Dependence}},
  \href{https://doi.org/10.1007/JHEP10(2013)087}{\emph{JHEP} {\bfseries 10}
  (2013) 087} [\href{https://arxiv.org/abs/1308.2627}{{\ttfamily 1308.2627}}].

\bibitem{Alonso:2013hga}
R.~Alonso, E.~E. Jenkins, A.~V. Manohar and M.~Trott, \emph{{Renormalization
  Group Evolution of the Standard Model Dimension Six Operators III: Gauge
  Coupling Dependence and Phenomenology}},
  \href{https://doi.org/10.1007/JHEP04(2014)159}{\emph{JHEP} {\bfseries 04}
  (2014) 159} [\href{https://arxiv.org/abs/1312.2014}{{\ttfamily 1312.2014}}].

\bibitem{Bijnens:1999sh}
J.~Bijnens, G.~Colangelo and G.~Ecker, \emph{{The Mesonic chiral Lagrangian of
  order $p^6$}},
  \href{https://doi.org/10.1088/1126-6708/1999/02/020}{\emph{JHEP} {\bfseries
  02} (1999) 020} [\href{https://arxiv.org/abs/hep-ph/9902437}{{\ttfamily
  hep-ph/9902437}}].

\bibitem{Weinberg:1978kz}
S.~Weinberg, \emph{{Phenomenological Lagrangians}},
  \href{https://doi.org/10.1016/0378-4371(79)90223-1}{\emph{Physica A}
  {\bfseries 96} (1979) 327}.

\bibitem{Webber:2008aa}
J.~Webber, \emph{Mesonic lagrangians and anomalous processes},  Master's
  thesis, {Mainz Institut f\"ur Kernphysik}, 2008.

\bibitem{Bijnens:2018lez}
J.~Bijnens, N.~Hermansson-Truedsson and S.~Wang, \emph{{The order $p^{8}$
  mesonic chiral Lagrangian}},
  \href{https://doi.org/10.1007/JHEP01(2019)102}{\emph{JHEP} {\bfseries 01}
  (2019) 102} [\href{https://arxiv.org/abs/1810.06834}{{\ttfamily
  1810.06834}}].

\bibitem{Graf:2020yxt}
L.~Graf, B.~Henning, X.~Lu, T.~Melia and H.~Murayama, \emph{{2, 12, 117, 1959,
  45171, 1170086, \textellipsis{}: a Hilbert series for the QCD chiral
  Lagrangian}}, \href{https://doi.org/10.1007/JHEP01(2021)142}{\emph{JHEP}
  {\bfseries 01} (2021) 142}
  [\href{https://arxiv.org/abs/2009.01239}{{\ttfamily 2009.01239}}].

\bibitem{Coleman:1969sm}
S.~R. Coleman, J.~Wess and B.~Zumino, \emph{{Structure of phenomenological
  Lagrangians. 1}}, \href{https://doi.org/10.1103/PhysRev.177.2239}{\emph{Phys.
  Rev.} {\bfseries 177} (1969) 2239}.

\bibitem{Callan:1969sn}
C.~G. Callan, S.~R. Coleman, J.~Wess and B.~Zumino, \emph{{Structure of
  phenomenological Lagrangians. 2}},
  \href{https://doi.org/10.1103/PhysRev.177.2247}{\emph{Phys. Rev.} {\bfseries
  177} (1969) 2247}.

\bibitem{DasBakshi:2022mwk}
S.~Das~Bakshi, M.~Chala, A.~D\'\i{}az-Carmona and G.~Guedes, \emph{{Towards the
  renormalisation of the Standard Model effective field theory to dimension
  eight: bosonic interactions II}},
  \href{https://doi.org/10.1140/epjp/s13360-022-03194-5}{\emph{Eur. Phys. J.
  Plus} {\bfseries 137} (2022) 973}
  [\href{https://arxiv.org/abs/2205.03301}{{\ttfamily 2205.03301}}].

\end{thebibliography}\endgroup

\end{document}